\documentclass[twocolumn,preprintnumbers,aps,prb,superscriptaddress,amsmath,amssymb]{revtex4-1}

\usepackage{graphicx}
\usepackage{dcolumn}
\usepackage{bm}
\usepackage{bbm}
\usepackage{hyperref}
\usepackage{color}
\usepackage{multirow}
\usepackage{physics}
\usepackage{xr}

\begin{document}

\title{Laser modulation of superconductivity in a cryogenic widefield nitrogen-vacancy microscope}

\author{Scott E. Lillie}
\affiliation{Centre for Quantum Computation and Communication Technology, School of Physics, The University of Melbourne, VIC 3010, Australia}
\affiliation{School of Physics, The University of Melbourne, VIC 3010, Australia}

\author{David A. Broadway}
\affiliation{Centre for Quantum Computation and Communication Technology, School of Physics, The University of Melbourne, VIC 3010, Australia}
\affiliation{School of Physics, The University of Melbourne, VIC 3010, Australia}

\author{Nikolai Dontschuk}
\affiliation{Centre for Quantum Computation and Communication Technology, School of Physics, The University of Melbourne, VIC 3010, Australia}
\affiliation{School of Physics, The University of Melbourne, VIC 3010, Australia}

\author{Sam C. Scholten}
\affiliation{Centre for Quantum Computation and Communication Technology, School of Physics, The University of Melbourne, VIC 3010, Australia}
\affiliation{School of Physics, The University of Melbourne, VIC 3010, Australia}

\author{Brett C. Johnson}
\affiliation{Centre for Quantum Computation and Communication Technology, School of Physics, The University of Melbourne, VIC 3010, Australia}
\affiliation{School of Physics, The University of Melbourne, VIC 3010, Australia}

\author{Sebastian Wolf}
\affiliation{School of Physics, The University of Melbourne, VIC 3010, Australia}

\author{Stephan Rachel}
\affiliation{School of Physics, The University of Melbourne, VIC 3010, Australia}

\author{Lloyd C. L. Hollenberg}
\email{Corresponding author: lloydch@unimelb.edu.au}
\affiliation{Centre for Quantum Computation and Communication Technology, School of Physics, The University of Melbourne, VIC 3010, Australia}
\affiliation{School of Physics, The University of Melbourne, VIC 3010, Australia}

\author{Jean-Philippe Tetienne}
\email{Corresponding author: jtetienne@unimelb.edu.au}
\affiliation{School of Physics, The University of Melbourne, VIC 3010, Australia}

\date{\today}

\begin{abstract}

Microscopic imaging based on nitrogen-vacancy (NV) centres in diamond, a tool increasingly used for room-temperature studies of condensed matter systems, has recently been extended to cryogenic conditions. However, it remains unclear whether the technique is viable for imaging temperature-sensitive phenomena below 10 K given the inherent laser illumination requirements, especially in a widefield configuration. Here we realise a widefield NV microscope with a field of view of $100$\,$\mu$m and a base temperature of $4$\,K, and use it to image Abrikosov vortices and transport currents in a superconducting Nb film. We observe the disappearance of vortices upon increase of laser power and their clustering about hot spots upon decrease, indicating that laser powers as low as $1$\,mW ($4$ orders of magnitude below the NV saturation) are sufficient to locally quench the superconductivity of the film ($T_c = 9$\,K). This significant local heating is confirmed by resistance measurements, which reveal the presence of large temperature gradients (several K) across the film. We then investigate the effect of such gradients on transport currents, where the current path is seen to correlate with the temperature profile even in the fully superconducting phase. In addition to highlighting the role of temperature inhomogeneities in superconductivity phenomena, this work establishes that, under sufficiently low laser power conditions, widefield NV microscopy enables imaging over mesoscopic scales down to 4 K with a submicrometer spatial resolution, providing a new platform for real-space investigations of a range of systems from topological insulators to van der Waals ferromagnets. 

\end{abstract}

\maketitle

Nitrogen-vacancy (NV) centre microscopy \cite{Doherty2013,Rondin2014} is a multi-modal imaging platform increasingly used to interrogate biological \cite{Schirhagl2014} and condensed matter systems\cite{Casola2018} at room temperature, where the long coherence times of the NV spin state enable high sensitivities in ambient conditions.\cite{Maze2008,Balasubramanian2008} Recent experimental efforts have extended NV microscopy to cryogenic temperatures,\cite{Kolkowitz2015,Thiel2016,Pelliccione2016} at which it can been used to interrogate low temperature phenomena such as transport and magnetism in low dimensional systems.\cite{Andersen2019,Thiel2019} Superconductivity is one such phenomenon to which NV microscopy is particularly applicable.\cite{Bouchard2011,Acosta2019} Previous studies have mainly focused on high-$T_c$ superconductors, probing the Meissner effect with ensembles of NV centres\cite{Waxman2014,Alfasi2016,Nusran2018,Joshi2019,Xu2019a} and achieving nanoscale imaging of microstructures and Abrikosov vortices using scanning single NV experiments.\cite{Thiel2016,Pelliccione2016,Rohner2018} Vortices in a high-$T_c$ superconductor have also been imaged using widefield imaging of NV ensembles.\cite{Schlussel2018} 

However, the viability of NV microscopy for more temperature-sensitive systems, such as superconductors with a $T_c\lesssim10$~K or electronic systems in the ballistic regime, remains to be seen. Indeed, at low temperatures the ``non-invasiveness" of the NV imaging platform becomes questionable, given the appreciable laser intensity impinging on the sample (up to $\sim1$~mW/$\mu$m$^2$, corresponding to saturation of the NV optical cycling) and microwave power (up to mWs) necessary to initialise, manipulate, and read out the NV spin state. Application of these fields can cause undesirable heating of the sample of interest and hence potentially affect the imaged phenomenon. Widefield imaging has many advantages over scanning single NV imaging, namely the ability to rapidly interrogate structures over large ($10$s$ - 100$s\,$\mu$m) fields of view\cite{Steinert2010,Pham2011} and perform multi-modal measurements simultaneously,\cite{Broadway2018b,Lillie2018,Broadway2019} however, it presents additional challenges owing the correspondingly large illumination area requiring total laser powers of up to $100$'s of mW. Such powers are much larger than the typical cooling power provided by helium bath or closed-cycle cryostats at a base temperature of 4 K, casting doubt on the possibility to operate at this temperature.

In this work, the impact of these essential components of NV microscopy is assessed by imaging superconducting niobium (Nb) devices in a cryostat with a base temperature ($\approx4$\,K) close to its critical temperature ($T_c \approx 9$\,K). Superconducting phenomena, namely the nucleation of Abrikosov vortices,\cite{Abrikosov1957} are used to assess the local heating from the excitation laser. We use electrical resistance measurements to quantify heating along the conduction path across the film, and compare to the insight from local imaging. Additionally, we image transport currents within a Nb device in both fully superconducting and normal states, and observe a non-uniform current distribution in the superconducting case which we associate with the non-uniform temperature profile due to the laser. This work has implications for future low temperature imaging experiments using NV microscopy, in both widefield and confocal configurations, identifying a regime of laser power conditions for which operation at sample temperature approaching 4 K is possible. Under these minimally-invasive conditions, the widefield NV microscope demonstrated here is an appealing tool for condensed matter studies, which may enable real-space investigations of a range of phenomena such as transport in topological insulators or in low-dimensional electronic systems, magnetisation dynamics in van der Waals ferromagnets and heterostructures, and superconductivity in 2D materials, to name just a few.   

\begin{figure}[h!]
	\begin{center}
		\includegraphics[width=1.0\columnwidth]{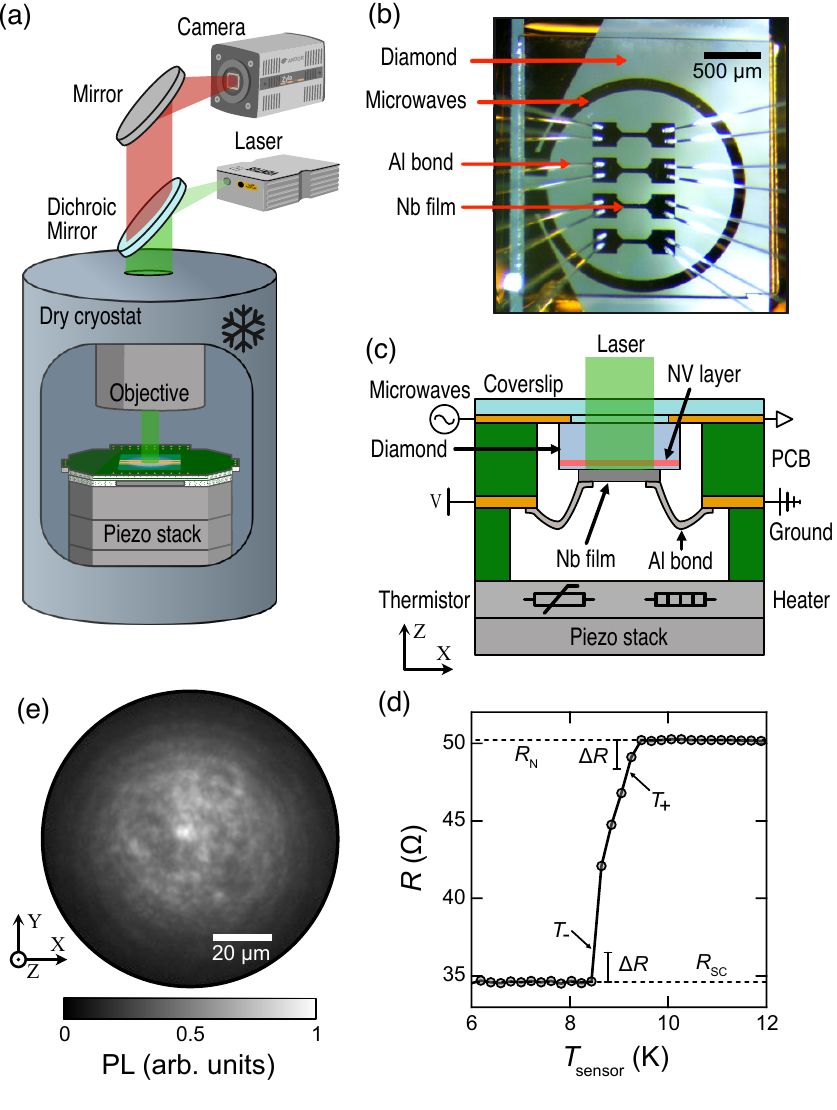}
		\caption{\textbf{A cryogenic widefield nitrogen-vacancy microscope:} (a) Schematic of the cryogenic widefield NV microscope (see description in text). (b) Photograph of thin-film Nb devices fabricated on the NV-diamond surface. The diamond is mounted to a cover slip with a microwave resonator visible beneath the diamond. (c) Schematic of the sample mount (see description in text). (d) $R$ versus $T_\text{sensor}$ for a single Nb device. The resistance drops from the normal state resistance, $R_\text{N}$, to the superconducting resistance, $R_\text{SC}$, and is characterised by temperatures $T_+$ and $T_-$ which differ from $R_\text{N}$ and $R_\text{SC}$ by $\Delta R = 0.05\times(R_\text{N} - R_\text{SC})$ respectively. (f) NV-layer PL image demonstrating the $100$\,$\mu$m field of view, taken under continuous wave (CW) laser illumination at $P_\text{laser} = 1.0$\,mW. Coordinate system used for all subsequent measurements is defined.}
		\label{Fig1}
	\end{center}
\end{figure}

\section*{Results}

The cryogenic widefield NV microscope used for this study houses the sample in a closed-cycle cryostat with a base temperature down to $4.0$\,K (see Methods for details), equipped with a superconducting vector magnet allowing application of a uniform magnetic field up to $1$\,T along an arbitrary axis, $\bf{B}_\text{app}$. The cryostat contains a cage mounted optics column which includes a high numerical aperture objective lens, and is accessible via a window at the top of the cryostat [Fig.\,\ref{Fig1}(a)]. A $532$\,nm laser is used to initialise and readout the NV-spin ensemble, and the red photoluminescence (PL) ($650-800$\,nm),\cite{Chen2011} is collected via the same optics column and focused onto an sCMOS camera [Fig.\,\ref{Fig1}(a)]. The NV-diamond used in these experiments is a 50-$\mu$m-thick membrane irradiated to form an NV imaging layer extending $200$\,nm below the diamond surface (see Methods).

Four $200$-nm-thick Nb devices were fabricated directly on the diamond surface, each device featuring two square bonding pads ($500$\,$\mu$m) connected by a narrow channel ($500$\,$\mu$m~$\times$~40~$\mu$m) [Fig.\,\ref{Fig1}(b)] (see Methods). The diamond was mounted to a glass cover slip featuring an omega-shaped microwave resonator for NV spin-state driving, which itself was mounted to a printed circuit board (PCB) for electrical contact to the resonator and Nb devices. This PCB was mounted to a stage equipped with a thermistor and heater for control and measurement of the near-sample temperature. Imaging of the near surface NV-layer occurs through the cover slip and bulk of the diamond, with the devices on the underside of the sample [Fig.\,\ref{Fig1}(c)].

Prior to imaging, the devices were characterised electrically to identify their critical temperature. The resistance across each device ($R$) was measured as a function of temperature as read by the thermistor on the sample holder ($T_{\text{sensor}}$). All devices showed a superconducting transition at a critical temperature $T_c \approx 9$\,K, where, for example, $R$ decreases from the normal state resistance, $R_\text{N} = 50.2$\,$\Omega$, to the superconducting resistance, $R_\text{SC} = 34.6$\,$\Omega$, which corresponds to the resistance of the non-superconducting leads [Fig.\,\ref{Fig1}(d)]. A typical PL image of the NV-layer beneath the Nb film under continuous wave (CW) illumination shows the Gaussian beam profile, with a $43$\,$\mu$m beam waist, giving reasonable illumination across a $100$\,$\mu$m field of view [Fig.\,\ref{Fig1}(e)]. At a total laser power of $P_\text{laser} = 1.0$\,mW, we note that imaging the sample results in only a slight change of the temperature as measured by the thermistor $\Delta T_\text{sensor} = 0.05$\,K.

\begin{figure*}[t]
	\begin{center}
		\includegraphics[width=1.0\textwidth]{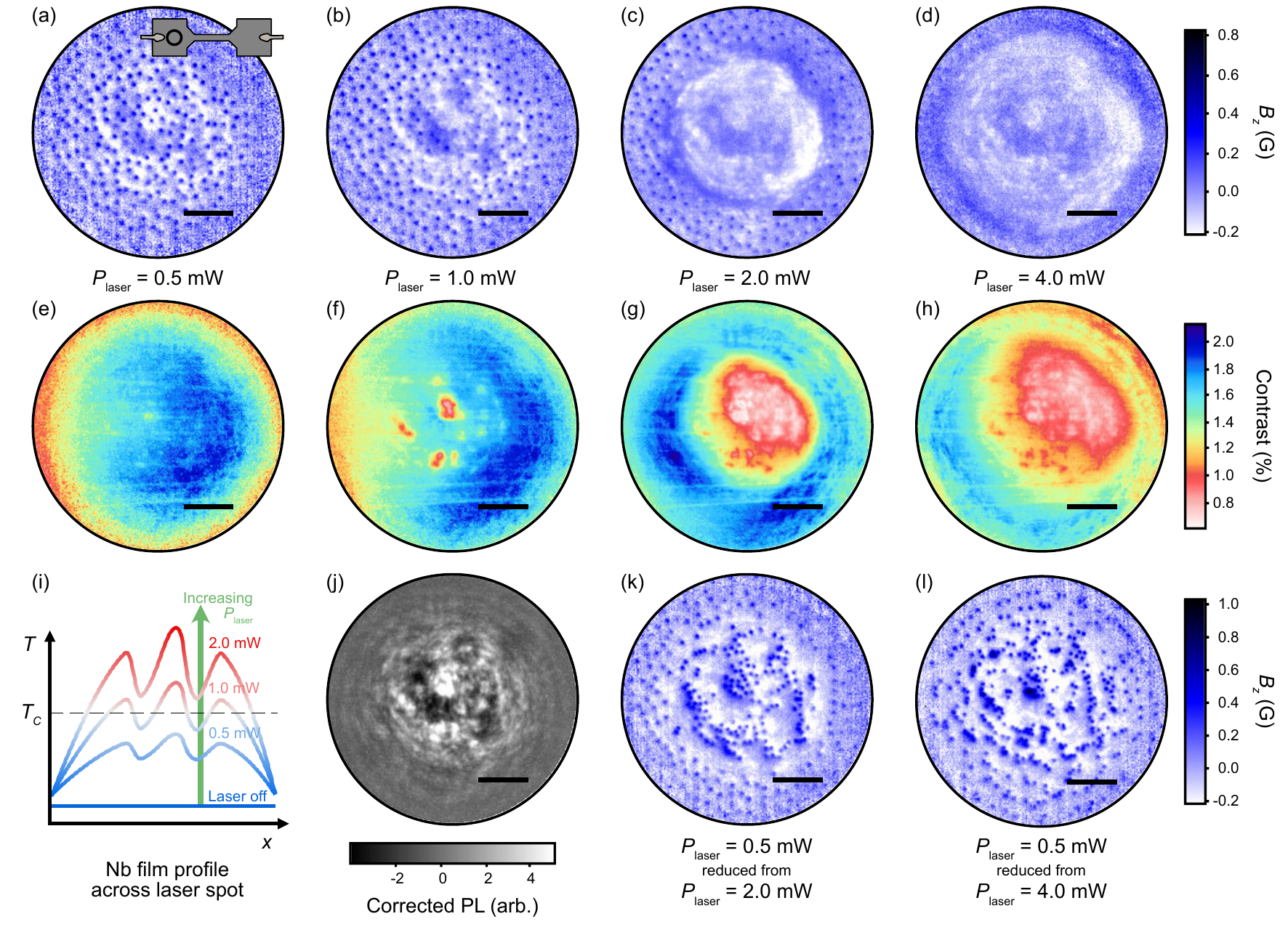}
		\caption{\textbf{Laser heating of Abrikosov vortices:} (a) - (d) $B_z$ images of vortices measured by ODMR at $P_\text{laser} = 0.5$, $1.0$, $2.0$, and $4.0$\,mW respectively. The location of the imaged region is indicated on the inset diagram in (a). The sample was cooled to base temperature ($T_\text{sensor} = 4.3$\,K) with the laser off, at a field $B_\text{app} = 1.5$\,G, and imaged at the same field. A background subtraction algorithm was applied to remove features varying over $20$\,pixel length scales or greater, which we attribute to artifacts in the frequency fitting at reduced contrast at low fields (see SI, section II). (e) - (h) Maps of the PL contrast of a single resonance line in the ODMR measurement used to reconstruct the $B_z$ images in (a) - (d). The reduced contrast near sites of vortex suppression indicates local reduction in the MW field strength (see SI, section II). (i) Illustration of laser heating of the Nb film at increasing $P_\text{laser}$, which is at base temperature when the laser is off. A temperature profile is imprinted on the Nb film by the laser, but remains below $T_c$ ($P_\text{laser} = 0.5$\,mW). Increasing $P_\text{laser}$ gives pockets of normal state Nb where the laser is most intense, removing the vortices ($P_\text{laser} = 1.0$\,mW). Further increasing $P_\text{laser}$ gives a large area of normal state Nb centred on the laser spot ($P_\text{laser} = 2.0$\,mW). (j) PL image highlighting the local variations in the laser beam profile across the Nb film. Broader variations in the PL (varying over length scales $20$ pixels or larger) have been subtracted to emphasise deviations from the approximately Gaussian profile. (k) and (l) $B_z$ images showing vortex clustering around hot spots when the laser power is reduced to $P_\text{laser} = 0.5$\,mW after imaging at $P_\text{laser} = 2.0$\,mW and $P_\text{laser} = 4.0$\,mW respectively. All scale bars are $20$\,$\mu$m.}
		\label{Fig2}
	\end{center}
\end{figure*}

We now move to NV magnetic imaging of a Nb device, focusing initially on vortices in the large Nb pad under no applied current. The sample was cooled to $T_\text{sensor} = 4.3$\,K under a uniform magnetic field perpendicular to the film plane ($z$ axis), $B_\text{app} = 1.5$\,G, with the laser off. The net magnetic field was imaged using CW optically detected magnetic resonance (ODMR), at a range of laser powers. The magnetic field images are presented as the field in the $z$ direction, $B_z$, which is calculated from projection of the field along the NV axes (see SI, section II). At $P_\text{laser} = 0.5$\,mW, the $B_z$ image shows vortices distributed across the field of view [Fig.\,\ref{Fig2}(a)], with a number density that is consistent with the theoretical value for such films, $n = B_\text{app}/\Phi_0$, where $\Phi_0$ is the magnetic flux quantum\cite{Stan2004} (see SI, Fig. S8). At $P_\text{laser} = 1.0$\,mW, vortices disappear from pockets near the centre of the image, while those towards the edge remain fixed [Fig.\,\ref{Fig2}(b)]. At $P_\text{laser} = 2.0$ and $4.0$\,mW, we see a disc centred on the laser spot in which the vortices are removed while the vortices towards the edge remain fixed [Figs.\,\ref{Fig2}(c,d)]. Comparing the $B_z$ images with ODMR contrast maps from the same measurements [Fig.\,\ref{Fig2}(e)-(h)], we observe that the regions where the vortices are removed correlate with regions in which the contrast is reduced, indicating a local reduction in the MW field strength (see SI, section II).

These observations are explained by local laser heating of the Nb film raising the temperature of some regions within the field of view above $T_c$ [Fig.\,\ref{Fig2}(i)]. As $P_\text{laser}$ is increased from $0.5$\,mW to $1.0$\,mW, pockets of normal state Nb are formed where the local laser intensity is largest (and $T > T_c$), thereby removing the vortices and attenuating the MW field (the superconducting state Nb is mostly transparent to the MW field given that the frequencies used are less than the superconducting gap). These regions are highlighted by subtracting the broader Gaussian curve from a PL image [Fig.\,\ref{Fig2}(j)]. Increasing $P_\text{laser}$ further, these pockets merge to form a normal state disc centred in the field of view.

Additionally, we observe clustering of vortices around intensity maxima in the laser profile when $P_\text{laser}$ is reduced from powers giving large areas of normal state Nb ($P_\text{laser} = 2.0$ and $4.0$\,mW) to a less invasive power ($P_\text{laser} = 0.5$\,mW) [Fig.\,\ref{Fig2}(k,l)]. This is because as the normal region shrinks, the vortices re-nucleate in the superconducting region where they are attracted by the nearest hot spot,\cite{Vadimov2018} i.e. here the normal/superconducting boundary. The vortices therefore cluster around local temperature maxima, where they are pinned once the Nb cools further (see SI, section III, for modelling and further discussion). A faster reduction in $P_\text{laser}$ reduces this clustering effect (see SI, Fig. S9). Moreover, uniform vortex configurations are recovered by heating the system globally above $T_c$, and cooling in the absence of laser (see SI, Fig. S9). Recently, thermal gradients arising from focused laser beams have achieved patterning at the single vortex level,\cite{Veshchunov2016} while ensembles of vortices can be manipulated by nano-patterned current profiles,\cite{Kalcheim2017} temperature patterning,\cite{Gonzalez2018} and local magnetic fields.\cite{Polshyn2019} 

\begin{figure*}[t]
	\begin{center}
		\includegraphics[width=1.0\textwidth]{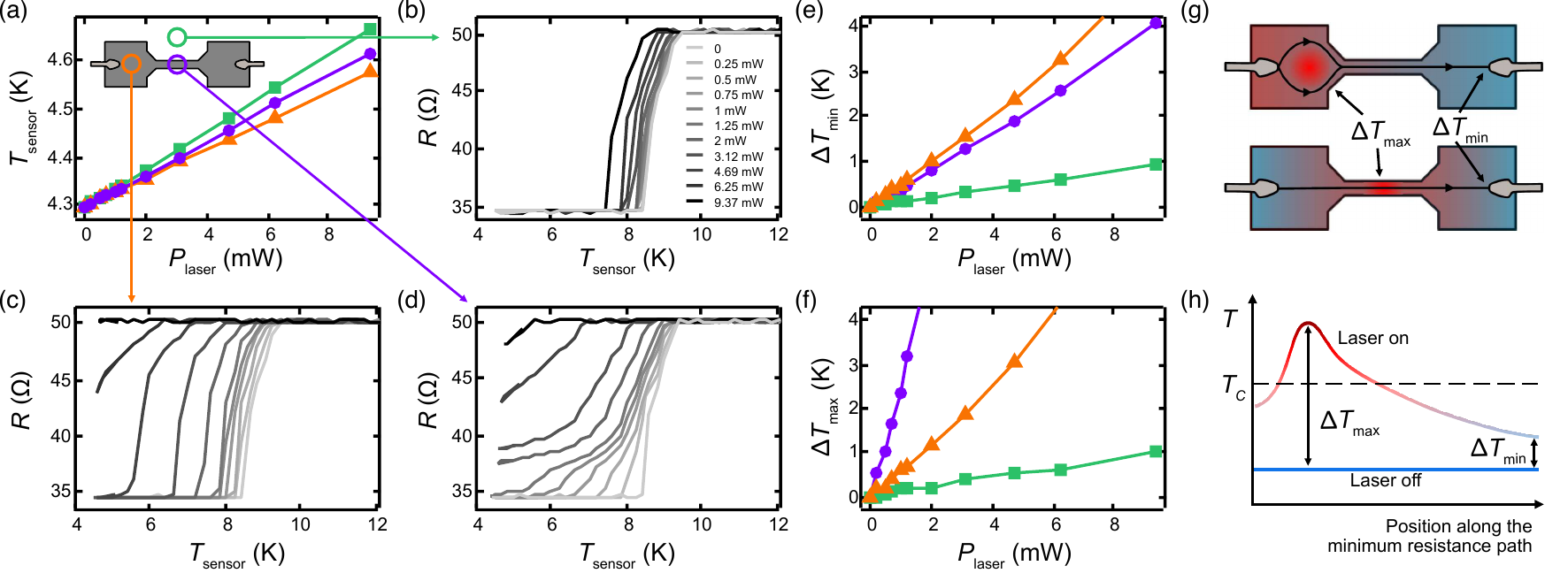}
		\caption{\textbf{Probing local laser heating with electrical measurements:} (a) $T_{\text{sensor}}$ versus $P_{\text{laser}}$ with no additional heating varying the location of the laser spot as identified on the inset diagram: $200$\,$\mu$m from the Nb film (orange), on the narrow Nb channel (purple), and on the centre of the Nb bonding pad (green). (b) - (d) $R$ versus $T_\text{sensor}$ for a single Nb device at increasing $P_{\text{laser}}$ (light to dark) with the laser spot focused on: (b) the diamond $200$\,$\mu$m from the Nb film; (c) the middle of the Nb bonding pad; and (d) the narrow Nb channel. The resistance was measured with a $10$\,$\mu$A current. (e) $\Delta T_\text{min}$ versus $P_\text{laser}$ for the laser spot focused at the identified locations. $\Delta T_\text{min}$ is the magnitude of the shift in $T_+$ from its value when the laser is off. (f) $\Delta T_\text{max}$ versus $P_{\text{laser}}$ for the laser spot locations. $\Delta T_\text{max}$ is the magnitude of the shift in $T_-$ from its value when the laser is off. (g) Schematic of two heating scenarios: one with laser focused on the bonding pad (upper); and the other with the laser focused on centre of narrow channel (lower). The location of $\Delta T_{\text{min}}$ and $\Delta T_{\text{max}}$ are indicated in each case. (h) Illustrative plot of the temperature distribution along the path of least resistance across the device with (red) and without (blue) laser heating, showing the location of $\Delta T_{\text{min}}$ and $\Delta T_{\text{max}}$.}
		\label{Fig3}
	\end{center}
\end{figure*}

The images presented demonstrate the invasiveness of widefield NV microscopy in this case. Imaging the Nb film with $P_\text{laser} = 2.0$\,mW gives a normal state region nearly the size of the laser spot, indicating heating in this region upwards of $> 5$\,K given the base temperature $4.3$\,K and the Nb $T_c \approx 9$\,K. Such a laser power corresponds to a relatively modest peak intensity of $40$\,W/cm$^2$, which is $4$ orders of magnitude lower than the laser intensity needed to saturate the optical cycling of the NV,\cite{Manson2006} as is often employed in single NV experiments. Surprisingly, the temperature at the sensor remains close to the base temp in these measurements, with $T_\text{sensor} < 4.4$\,K at $P_\text{laser} = 2$\,mW. A linear response of $T_\text{sensor}$ to $P_\text{laser}$ is observed [Fig.\,\ref{Fig3}(a)], with a slope that depends on the location of the laser beam: the thermistor is heated most efficiently when a greater portion of the laser is incident on the transparent diamond. In any case, $T_\text{sensor}$ remains below $4.7$\,K for $P_\text{laser}$ up to $10$\,mW, implying that there exist strong temperature gradients across the sample, which may be overlooked if local indicators are not available. 

To further quantify the local heating, we measure the resistance, $R$, as a function of $T_\text{sensor}$ (controlled by the heater) and $P_\text{laser}$, when the laser spot is focused at various locations around the device. When the laser is focused $200$\,$\mu$m from the Nb film, the $R$ versus $T_\text{sensor}$ curves shift to lower $T_\text{sensor}$ as $P_\text{laser}$ increases, indicating a global heating of the Nb device [Fig.\,\ref{Fig3}(b)]. When the laser spot is focused on the Nb bonding pad, the curves shift again, but the shape is severely distorted, as $T_-$ moves to lower $T_\text{sensor}$ at a faster rate than $T_+$ as $P_\text{laser}$ increases [Fig.\,\ref{Fig3}(c)] ($T_{+/-}$ defined in Fig.\,\ref{Fig1}(d)). This scenario is exacerbated when the laser is focused on the narrow channel, where the full width of the current path is encompassed by the laser spot [Fig.\,\ref{Fig3}(d)].

The minimum and maximum temperature increase along the current path through the device ($\Delta T_\text{min}$ and $\Delta T_\text{max}$) can be quantified by the shift in $T_+$ and $T_-$ at a given $P_\text{laser}$ from their values when the laser is off [Figs.\,\ref{Fig3}(e,f)]. In the case where the laser is focused on the narrow channel, where the current must pass through the maximally heated part of the device [Fig.\,\ref{Fig3}(g) lower], we find a maximum temperature increase of $> 2$\,K/mW, sufficient to completely quench the superconductivity of the illuminated section of the strip at $P_\text{laser} = 2$\,mW. This is consistent with Fig.\,\ref{Fig2}(g) where imaging at $P_\text{laser} = 2.0$\,mW results in a normal state Nb disc centred on the laser spot $\sim 50$\,$\mu$m in diameter. The minimum temperature increase of $\sim 0.5$\,K/mW suggests that the whole Nb device experiences significant heating even $1$\,mm away from the laser spot. This is possibly indicative of a relatively poor thermal conductivity of our implanted diamond substrate, and/or a poor thermal contact with the Nb film. When the laser is focused on the bonding pad, the current can avoid the hottest part of the device directly under the laser spot [Fig.\,\ref{Fig3}(g) upper], and so $\Delta T_\text{max}$ is reduced and is closer to the global minimum heating of $\sim 0.5$\,K/mW as compared to the previous case [Fig.\,\ref{Fig3}(h)]. The heating of the Nb film is reduced but still measurable with the
laser off of the Nb device ($\sim0.2$\,mm away), $\sim 0.1$\,K/mW, just a factor $3$ shy of the heating measured at the sensor. Heating of the sample by the microwave field was also assessed and found to give a small global temperature change ($\sim 0.1$\,K) across the device at powers relevant to most imaging applications (see SI, section I).

\begin{figure*}[t]
	\begin{center}
		\includegraphics[width=1.0\textwidth]{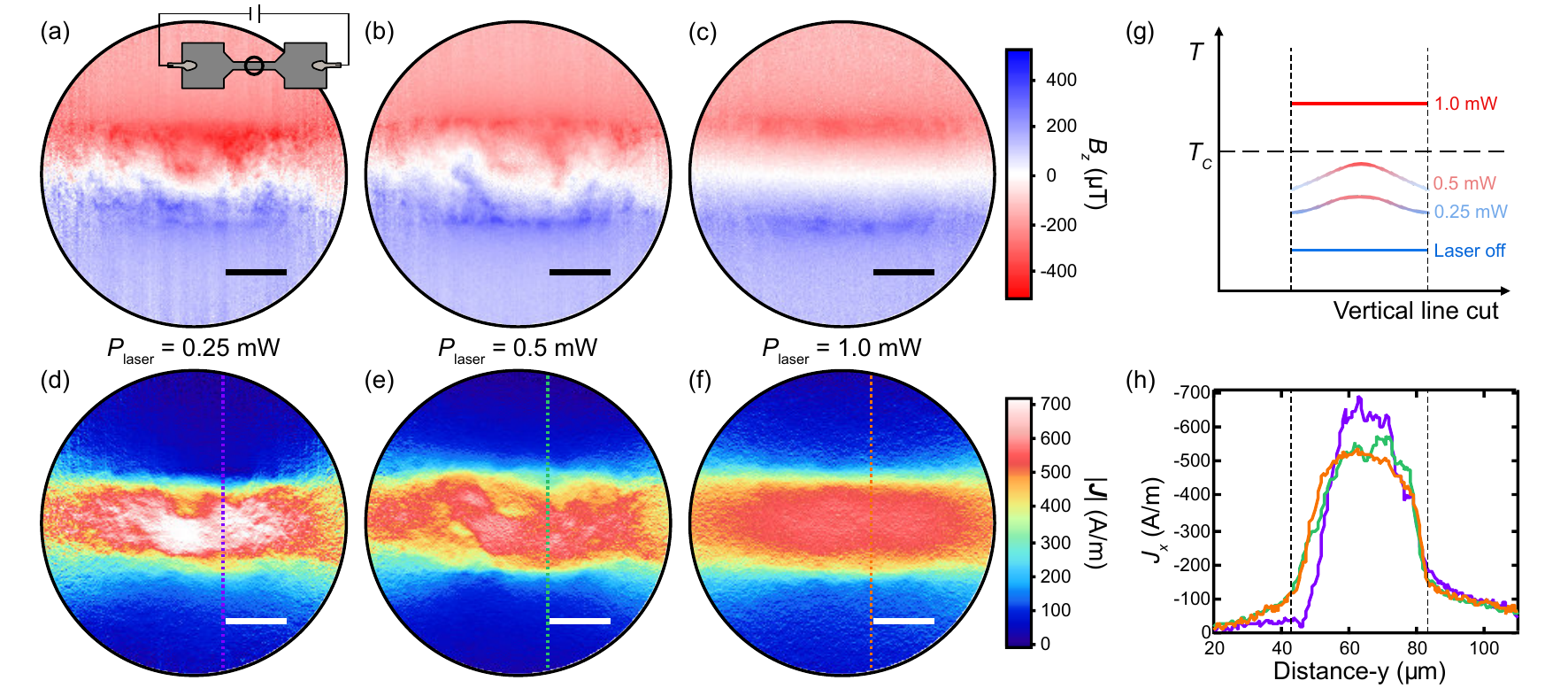}
		\caption{\textbf{Laser heating of superconducting transport:} (a) - (c) Images of the magnetic field in the z-direction due to charge transport within the Nb channel, imaged at $P_\text{laser} = 0.25$\,mW, $0.5$\,mW, and $1.0$\,mW. The background field that facilitates the measurement has been subtracted. (d) - (f) Total current density map reconstructed from the accompanying $B_z$ images (see SI, section IV, for reconstruction details). A total current of $20$\,mA was maintained throughout the acquisition of these images. The device resistance when imaging with $P_\text{laser} = 0.25$\,mW and $0.5$\,mW was $R = 34.6$\,$\Omega$, indicating a fully superconducting current path through the device, whereas for $P_\text{laser} = 1.0$\,mW, the resistance was $R = 50.2$\,$\Omega$, indicating a normal state Nb film. All scale bars are $20$\,$\mu$m. (g) Illustration of the temperature profile across the Nb channel at each of the laser conditions imaged. At $P_\text{laser} = 0.25$ and $0.5$\,mW the laser heats the channel reducing the critical current density as $T$ approaches $T_c$. At $P_\text{laser} = 1.0$\,mW, Joule heating from the normal state Nb dominates and gives a near-uniform temperature profile. (h) Line cuts of $x$ component of the current density, $J_x$, across the same section of the Nb channel for $P_\text{laser} = 0.25$\,mW (purple), $0.5$\,mW (green), and $1.0$\,mW (orange). The location of the line cuts are indicated in (d-f). The edges of the Nb channel, indicated by vertical dashed lines, were identified from the ODMR contrast (see SI, Fig. S11).}
		\label{Fig4}
	\end{center}
\end{figure*}

Resistance measurements allow us to characterise local heating due to the laser and infer its impact on the current path, however, the impact of laser heating on the local current distribution can be imaged directly by measuring the \O rsted field using ODMR. Direct reconstruction of current paths in superconductors is of particular interest to superconducting single-photon and single-electron detectors, which rely on quenching superconductivity at the site of detection.\cite{Bulaevskii2012,Adami2013} Here we apply a biasing field, $\bf{B}\rm_\text{app} = (47.5,97.4,19.1)$\,G, to resolve the different NV orientations and measure the net vector magnetic field, $\bold{B} = (B_x$,$B_y$,$B_z)$. This allows us to reconstruct the current density in the Nb device with good accuracy, by inverting the Biot-Savart law\cite{Tetienne2017,Tetienne2019} (see SI, section IV). The \O rsted field was measured at three different laser powers, $P_\text{laser} = 0.25$, $0.5$, and $1.0$\,mW, with a constant total current, $I = 20$\,mA, for all measurements. We show only the field projected in the $z$ direction, $B_z$, as this was used to reconstruct the two-dimensional current density, $\bf{J}$ (see SI, section IV).

At $P_\text{laser} = 0.25$\,mW and $0.5$\,mW, $R = 34.6$\,$\Omega$ for the duration of measurement (approximately $1$\,hour), indicating fully superconducting current pathways across the device (see Methods). The measured $B_z$ [Fig.\,\ref{Fig4}(a,b)] and the associated total current density, $|\bf{J}|$ [Fig.\,\ref{Fig4}(d,e)], show a non-uniform current density distribution, that is more laterally confined at the lower laser power. At $P_\text{laser} = 1.0$\,mW, $R = 50.2$\,$\Omega$, indicating a fully normal state current pathway, as expected from cascade Joule heating from the comparably large current. Consequently, $B_z$ [Fig.\,\ref{Fig4}(c)] and $|\bf{J}|$ [Fig.\,\ref{Fig4}(f)] show a current distribution that is uniform across nearly the full width of the Nb channel. The current density tapers at the edge of the field of view due to a reconstruction error (see SI, Fig. S14, where we show that $|\bf{J}|$ reconstructed from $B_y$ does not show this tapering effect). Note that imaging with $P_\text{laser} = 0.5$\,mW but using the heater to raise the temperature above $T_c$ ($T_{\rm sensor}=12$\,K) gave results identical to the $P_\text{laser} = 1.0$\,mW case, for both the $|\bf{J}|$ map and the ODMR contrast.

The non-uniform current density through the superconducting state Nb channel is a direct consequence of the temperature profile imprinted by the laser [Fig.\,\ref{Fig4}(g)]. As the local temperature increases, the superconducting gap, and hence the critical current density, $J_c$, is reduced. The measured current density is therefore larger where the local temperature is lower. Increasing the total laser power reduces $J_c$ across the channel, and the current density distribution broadens to maintain the same total current. Line cuts of the current density in the $x$ direction highlight this effect [Fig.\,\ref{Fig4}(h)]. We note that the imaged $|\bf{J}|$ under fixed laser conditions retains a consistent shape across the Nb channel as the field of view is translated, indicating that the non-uniformity observed arises from the excitation laser, rather than local variation in the film (see SI, Fig. S15).

\section*{Conclusion}

In this work, we demonstrated a cryogenic widefield NV microscope featuring a submicrometer spatial resolution and a field of view of $100~\mu$m, and applied it to the imaging of vortices and transport currents in superconducting Nb devices. The demonstrated field of view is five times larger than in the previous demonstration of widefield NV imaging at cryogenic temperatures, which reported a field of view of $20~\mu$m.\cite{Schlussel2018} This new capability is ideal for real-space investigations of mesoscopic phenomena in a variety of materials and devices, and also enables imaging of several samples in parallel. For instance, atomically thin samples of van der Waals materials prepared by mechanical exfoliation typically come in the form of multiple micrometre-sized flakes with different properties (thickness, shape), and so widefield imaging of such samples would allow simultaneous studies of many of them, greatly speeding up the characterisation process. This could be particularly useful to investigate, for example, the magnetic properties of ferromagnetic van der Waals materials and heterostructures. 

Our work also highlighted a limitation of NV microscopy for low temperature measurements, where the laser illumination required to optically interrogate the NV centres can lead to significant heating of the sample under study. Using the Nb superconducting film ($T_c\approx9$~K) as a local temperature probe, we found that even modest illumination powers (2 mW, corresponding to a peak intensity of $40$\,W/cm$^2$) can locally quench the superconductivity of the film, implying that the sample temperature exceeds 9 K even when the temperature measured with a nearby sensor remains below 5 K, close to the base temperature of the cryostat. This work thus demonstrates the need for caution in NV sensing experiments at low temperature, setting a limit to the laser power that can be used for minimally-invasive imaging. In our experiments, a total power of 0.5 mW (peak intensity of $10$\,W/cm$^2$) was sufficiently low to keep the Nb devices fully superconducting, allowing an array of frozen superconducting vortices to be imaged. Although the acceptable illumination conditions will depend on the details of the experimental setup and sample under study, it is likely that the most sensitive samples will require strategies to mitigate laser-induced heating. These include the introduction of a thin high-reflectance metallic film between the diamond and sample of interest, the use of better thermal conductors between the sample and the cooling elements of the cryostat, or the implementation of an optimised illumination geometry to reduce the required laser power. These precautions may be necessary even in single NV experiments which can have similar laser power densities at the point of imaging despite using less total power. Such steps will unlock the potential of NV microscopy for a broader range of low temperature condensed matter phenomena.

\section*{Methods}

\subsection*{Diamond samples}

The NV-diamond sample used in these experiments was made from a $4.4$\,mm~$\times$~$4.4$\,mm~$\times$~$50$\,$\mu$m type-Ib, single-crystal diamond substrate grown by high-pressure, high-temperature synthesis, with $\{100\}$-oriented polished faces (best surface roughness $<5$\,nm Ra), purchased from Delaware Diamond Knives. The diamond had an initial nitrogen concentration of [N]~$\sim 100$\,ppm. To create vacancies, the received plate was irradiated with $^{12}$C$^-$ ions accelerated at $100$\,keV with a dose of $10^{12}$ ions/cm$^2$. We performed full cascade Stopping and Range of Ions in Matter (SRIM) Monte Carlo simulations to estimate the depth distribution of the created vacancies [Fig.~S1], predicting a distribution spanning the range $0$~-~$200$\,nm with a peak vacancy density of $\sim110$~ppm at a depth of $\sim 130$\,nm. Following irradiation the diamond was laser cut into smaller $2.2$\,mm~$\times$~$2.2$\,mm~$\times$~$50$\,$\mu$m plates, which were then annealed in a vacuum of $\sim10^{-5}$~Torr to form the NV centres, using the following sequence:\cite{Tetienne2018} $6$\,h at $400$\,$^\circ$C, $6$\,h ramp to $800$\,$^\circ$C, $6$\,h at $800$\,$^\circ$C, $6$\,h ramp to $1100$\,$^\circ$C, $2$\,h at $1100$\,$^\circ$C, $2$\,h ramp to room temperature. After annealing the plates were acid cleaned ($15$\,minutes in a boiling mixture of sulphuric acid and sodium nitrate).

\subsection*{Nb device fabrication}

The devices were fabricated by e-beam evaporation of $200$\,nm of Nb through an Invar shadow mask. To achieve superconducting Nb, care was taken in preparing the high vacuum of our Thermionics e-beam evaporator. Namely, Nb pellets in a Fabmate crucible were first heated to evaporating temperatures with the sample shutter closed for $10$\,min. Then Ti was evaporated, again with the sample shutter closed, to bring the chamber vacuum below $10^{-7}$\,mbar via a sublimation effect. Finally the Nb was heated to evaporation temperature and the shutter was opened with an evaporation rate of $6$\,-\,$8$\,\AA/s, with chamber pressures below $2\times10^{-5}$\,mbar during evaporation. Care was taken to ensure sufficient Nb was present in the crucible to avoid damage from the high Nb evaporation temperatures, as visible crucible burning damage correlated with poor quality superconductivity (i.e., low $T_c$).

After fabrication of the Nb devices, the diamond was glued on a glass cover-slip, itself glued to a PCB, and the Nb devices were electrically connected to the PCB via Al wire bonds visible in Fig.\,\ref{Fig1}(b) of the main text. The total resistance at the base temperature was similar for all four devices, $R_\text{SC}\approx 35~\Omega$ given by the non-superconducting leads, however the resistance increase upon heating above $T_c$ varied across the devices (by a factor up to $2$) in correlation with the value of $T_c$. For all the experiments reported in the paper, we used the device with the highest $T_c\approx9$~K, which had a resistance in the normal state $R_\text{N}\approx50~\Omega$. From the difference $R_\text{N}-R_\text{SC}\approx15~\Omega$, we can estimate the resistivity of the Nb film, $\rho\approx6\times10^{-7}~\Omega$~m where we took the dimensions of the Nb strip to be $200/40/0.2$ (length/width/thickness in $\mu$m). Knowing that the product $\rho l$ for Nb is $\rho l=3.75\times10^{−16}~\Omega$~m$^2$ \cite{Ashcroft}, we deduce the mean free path for our Nb film, $l\approx0.6$~nm. For pure Nb, the coherence length is $\xi_0=38$~nm, hence we are in the dirty limit $l\ll \xi_0$. We can then use the dirty limit approximations to estimate the effective coherence length at zero temperature, $\xi(0)=0.85\sqrt{\xi_0 l}\approx4$~nm, and the magnetic penetration depth at zero temperature, $\lambda(0)=0.62 \lambda_L\sqrt{\xi_0/l}\approx200$~nm using the London penetration depth for pure Nb, $\lambda_L=39$~nm. At finite temperature, we have $\xi(T)=\xi(0)/\sqrt{1-T/T_c}$ and $\lambda(T)=\lambda(0)/\sqrt{1-T/T_c}$. At the base temperature of $T_{\rm base}=4.3$~K, this gives $\xi(T_{\rm base})\approx6$~nm and $\lambda(T_{\rm base})\approx300$~nm, but with laser heating these parameters can be significantly larger up to the point where the superconductivity is quenched as demonstrated in Fig.\,\ref{Fig2} of the main text.

\subsection*{Experimental setup}

The NV imaging was facilitated by a custom-built widefield fluorescence microscope, similar to those described in Refs.\cite{Simpson2016,Tetienne2017} built around a closed-cycle cryostat (Attocube attoDRY1000) equipped with a 1-T superconducting vector magnet (Cryomagnetics). Optical excitation is achieved by using a $532$\,nm continuous wave (CW) laser (Laser Quantum Ventus $1$\,W, coupled to a single-mode fiber), with pulsing enabled by a fibre-coupled acousto-optic modulator (AAOpto MQ180-G9-Fio). The laser is reflected into the optical column of the cryostat by a dichroic beam splitter, where a widefield lens ($f = 300$\,mm) focuses the beam on the back of a microscope objective (Attocube LT-APO/VISIR/0.82), controlling the laser spot size at the sample. The NV photoluminescence (PL) is collected along the same optical path, which includes two $250$\,mm lenses in the column in a $4f$ configuration to increase the field of view. The PL is separated from the excitation laser by the dichroic beam splitter, filtered through a $731/137$\,nm band pass filter, and imaged by focusing with a $300$\,mm tube lens onto a water cooled sCMOS camera (Andor Zyla 5.5-W USB3) for imaging. The microwave infrastructure used to drive NV spin-state transitions comprises a signal generator (Rohde \& Schwarz SMB100A), a switch (Mini-Circuits ZASWA-2-50DR+), and a $50$\,W amplifier (Mini-Circuits HPA-50W-63), which is passed to the microwave resonator via the custom made printed circuit board (PCB) which hosts the sample. Electrical control of the Nb devices was enacted by a source-measurement unit (Keithley SMU 2450), connected to the PCB to which the Nb devices were wire bonded. All measurements were sequenced using a pulse pattern generator (SpinCore PulseBlasterESR-PRO 500 MHz) to gate the laser and MW, and synchronize the image acquisition.

The cryostat integrates a superconducting vector magnet capable of applying up to 1 T in any direction. The cold plate of the cryostat is thermally coupled to the sample-holding optical column via a He exchange gas, the pressure of which tunes the base temperature at the sample down to $4.0$\,K. In these experiments, we use an exchange gas pressure such that $T_\text{base} = 4.3$\,K. The calculated optical resolution limit of the microscope is $\approx0.4$\,$\mu$m, given the objective numerical aperture of $0.82$ and target PL wavelengths $650-800$\,nm. In practice, the smallest resolvable features observed are down to $0.7$\,$\mu$m, as shown in an ODMR frequency shift map of stress features on the bare diamond surface [Fig.\,S2(a)], and quantified by fitting a line profile of the image [Fig.\,S2(b)]. This practical limit is likely due to optical aberrations, especially due to imaging through the $150$\,$\mu$m thick cover slip, and the $50$\,$\mu$m thick diamond.

\subsection*{Resistance measurements}

The resistance vs temperature (RT) measurements shown in Fig.\,\ref{Fig3} of the main text were taken with a Keithley 2450 Source Measure Unit operated in current source mode with a DC current of $I=10\,\mu$A. The temperature was controlled with a Lakeshore 335 Temperature Controller connected to a resistive heater and a calibrated sensor (Lakeshore Cernox CX-1050-CU-HT-1.4L), both thermally attached to the top Ti plate of the piezo stack. To obtain an RT curve, the temperature setpoint was incremented by $0.2$\,K in the increasing direction starting from the base temperature, and a series of measurements of both the resistance and temperature was recorded as the temperature settled. The number of measurements was chosen such that the temperature is mostly flat at the end of the series -- $20$ measurements over the course of $2$\,seconds in this case. The resistance and temperature of the final measurement of the series for each set-point were then combined to form an RT plot.

In Fig.\,4 of the main text, a DC current of $I=20$\,mA was applied during the ODMR measurement to allow the \O rsted magnetic field to be imaged. To characterise the effect of this current on the superconductivity of the Nb wire, we recorded RT sweeps at increasing current and laser powers, with the laser centred on the narrow section of the wire as in Fig.\,4 of the main text [Fig.\,S3]. Without laser illumination [Fig.\,S3(a)], the threshold $T_-$ is lowered to $T_-=7.4$\,K (a $1$\,K drop) with a current of $20$\,mA, which corresponds to a decrease in $T_c$ due to the relationship between current and $T_c$. On the other hand, $T_+$ changes by a larger amount and the transition becomes nearly vertical, but this is mainly because of Joule heating at this large current: when a small section of the Nb wire turns normal with a resistance of, e.g., $R=1\,\Omega$, there is a dissipated power $RI^2=0.4$\,mW in this section which will raise the temperature of the neighbouring sections, eventually quenching the superconductivity of the entire device. At $P_\text{laser}=0.25$\,mW and $0.5$\,mW [Fig.\,S3(b) and (c)], $T_-$ with $I=20$\,mA becomes $6.$\,K and $5.4$\,K, respectively, indicating a local temperature increase caused by the laser of $0.8$\,K and $2.0$\,K, respectively. Thus, at the base temperature (i.e. in the absence of active heating from the heater), the reduced temperature $t=T/T_c$ in this region is $t=0.70$ and $t=0.86$, assuming $T_c=7.4$K. At $P_\text{laser}=0.75$\,mW [Fig.\,S3(d)], the device is no longer superconducting at the base temperature with a current of $20$\,mA.

\section*{Acknowledgments}

This work was supported by the Australian Research Council (ARC) through Grants No. DE170100129, CE170100012, LE180100037, FT180100211, DP190101506 and DP200101118. We acknowledge the AFAiiR node of the NCRIS Heavy Ion Capability for access to ion-implantation facilities. D.A.B. and S.E.L. are supported by an Australian Government Research Training Program Scholarship. We acknowledge the contributions of D. Creedon and S. Yianni in developing the fabrication of superconducting Nb films, and A. Martin for useful discussions.

\bibliographystyle{apsrev4-1}
\bibliography{Superconductivity}

\renewcommand{\thefigure}{S\arabic{figure}}
\setcounter{figure}{0}
\setcounter{figure}{0}

\clearpage

\section*{Supporting Information}

\begin{figure}[ht]
	\begin{center}
		\includegraphics[width=1.0\columnwidth]{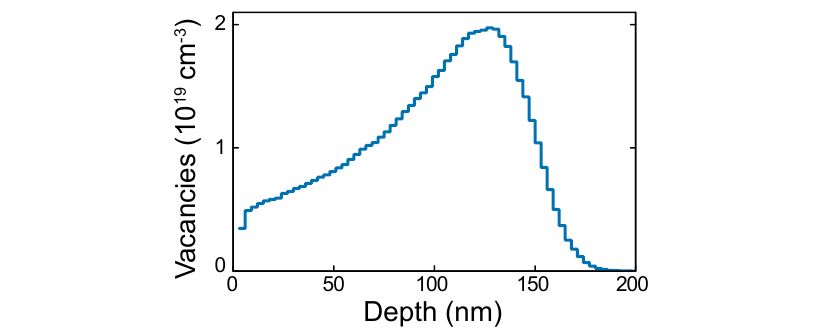}
		\caption{\textbf{Stopping and Range of Ions in Matter (SRIM) simulations:} Vacancy concentration as a function of depth for a $100$\,keV $^{12}$C$^-$ implant in diamond at a dose of $10^{12}$\,ions/cm$^2$. We assumed a diamond density of $3.51$\,g\,cm$^{-3}$ and a displacement energy of $50$\,eV. The peak vacancy concentration of $\sim2\times10^{19}$~cm$^{-3}$ corresponds to $\sim110$~ppm.}
		\label{FigSI_SRIM}
	\end{center}
\end{figure}

\begin{figure}[ht]
	\begin{center}
		\includegraphics[width=1\columnwidth]{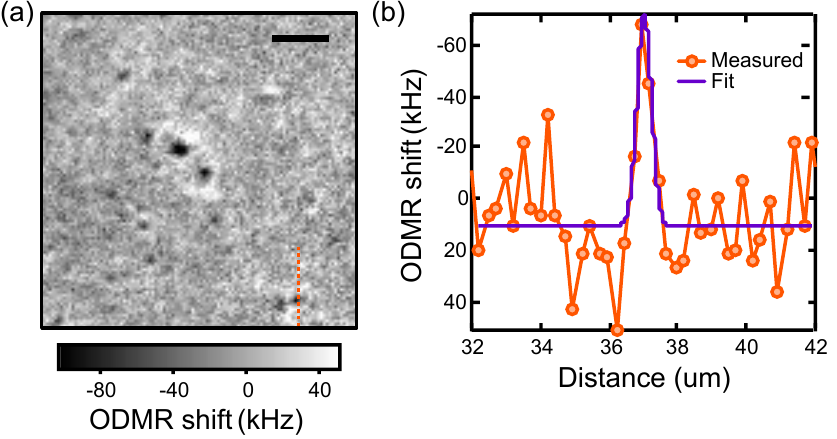}
		\caption{\textbf{Microscope optical resolution}: (a) ODMR frequency shift map (i.e. single NV spin-resonance frequency fit with a mean value subtraction) revealing stress features on the bare diamond surface, away from the Nb film. The scale bar is $5$\,$\mu$m. (b) ODMR shift profile across a small feature seen in (a) (orange), and Lorentzian fit (purple) showing a FWHM of $0.7$\,$\mu$m.}
		\label{FigSI_optical_res}
	\end{center}
\end{figure}

\begin{figure}[ht]
	\begin{center}
		\includegraphics[width=1.0\columnwidth]{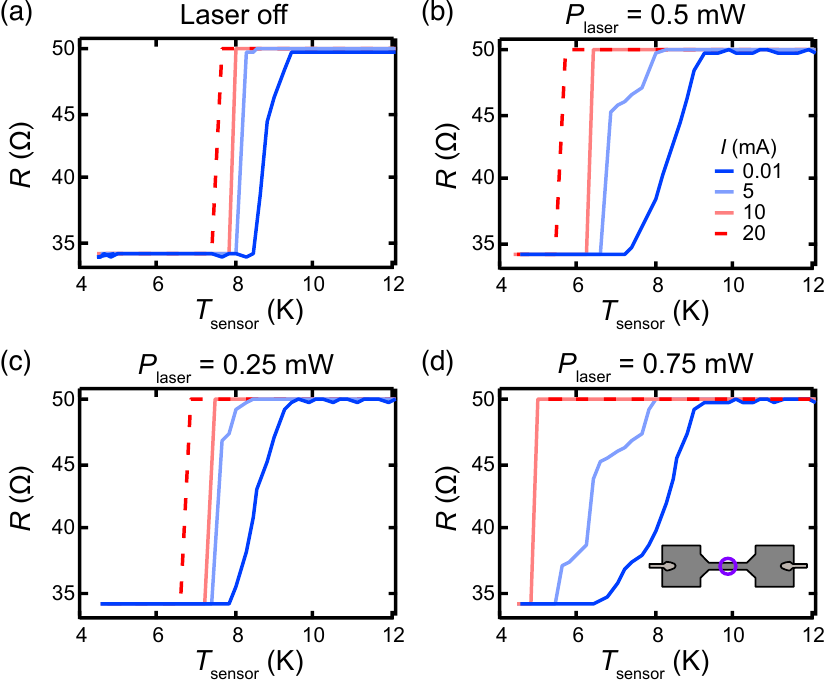}
		\caption{\textbf{RT curves vs current:} RT curves recorded with current values ranging from $I = 0.01$\,mA to $20$\,mA, and at different laser powers: (a) $P_\text{laser}=0$\,mW, (b) $0.25$\,mW, (c) $0.5$\,mW, and (d) $0.75$\,mW. The inset depicts the position of the laser spot (purple circle) for these measurements.}
		\label{FigSIRTvsI}
	\end{center}
\end{figure}

\section{Microwave heating}\label{MW}

\begin{figure}[ht]
	\begin{center}
		\includegraphics[width=1\columnwidth]{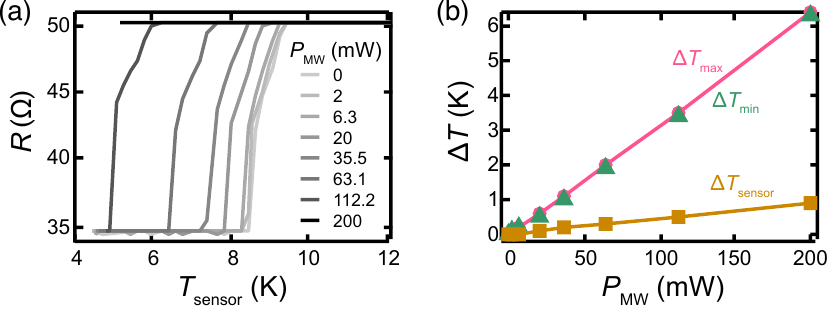}
		\caption{\textbf{Probing microwave heating with electrical measurements:} (a) $R$ versus $T_\text{sensor}$ with $P_\text{MW}$ ranging from $0$\,mW to $200$\,mW. The measurements were taken with no laser illumination. (b) Changes in temperature versus $P_\text{MW}$. The change in temperature at the sample is taken from the resistance measurements in (a), quantified by $\Delta T_\text{max}$ (pink) and $\Delta T_\text{min}$ (green), which are defined in the main text. The change in temperature $\sim2$\,mm below the sample is given by the change in thermistor reading from the case where the microwave field is off, $\Delta T_\text{sensor}$ (gold).}
		\label{FigSI_mw_heating}
	\end{center}
\end{figure}

Most NV experiments utilise a microwave resonator or strip-line to drive NV spin-state transitions. The required microwave field power varies greatly depending on the measurement, and the microwave-source-sample configuration. Here, we assess heating of the Nb sample due to the microwave resonator by measuring the device resistance and $T_\text{sensor}$ across a range of microwave powers, $P_\text{MW}$. The $R$ versus $T_\text{sensor}$ curves of the device imaged in the main text at various $P_\text{MW}$ show a translation of the critical temperature to lower $T_\text{sensor}$ as $P_\text{MW}$ is increased [Fig.\,\ref{FigSI_mw_heating}(a)]. The shape of the curve is preserved across the range of $P_\text{MW}$, as shown by matching changes in $\Delta T_\text{min}$ and $\Delta T_\text{max}$ (defined in main text) with $P_\text{MW}$ ($\sim32$\,mK/mW) [Fig.\,\ref{FigSI_mw_heating}(b)], indicating negligible temperature gradients across the length of the Nb film. The change in temperature as measured by the thermistor, $\Delta T_\text{sensor}$ is comparably small ($\sim 5$\,mK/mW), indicating a significant temperature gradient between resonator-sample plane, and the thermistor, which are separated by $>2$\,mm. The ODMR based imaging presented in the main text uses $P_\text{MW} = 2$\,mW, and hence, the microwave heating of the Nb film is negligible as compared to the laser heating in the same measurement.

\section{Vortex imaging}\label{lowfieldODMR}

The magnetic images of vortices presented in Fig.\,2 of the main text originate from continuous wave (CW) ODMR measurements, at low field, such that the vortex density is compatible with our imaging resolution limit, $\sim700$\,nm. Here, we describe these measurements in detail and the subsequent analysis. We show additional vortex images that demonstrate the dependence of vortex density on the magnetic field, and the dependence of vortex clustering on the cooling protocol used. 

\begin{figure*}[t]
	\begin{center}
		\includegraphics[width=0.8\textwidth]{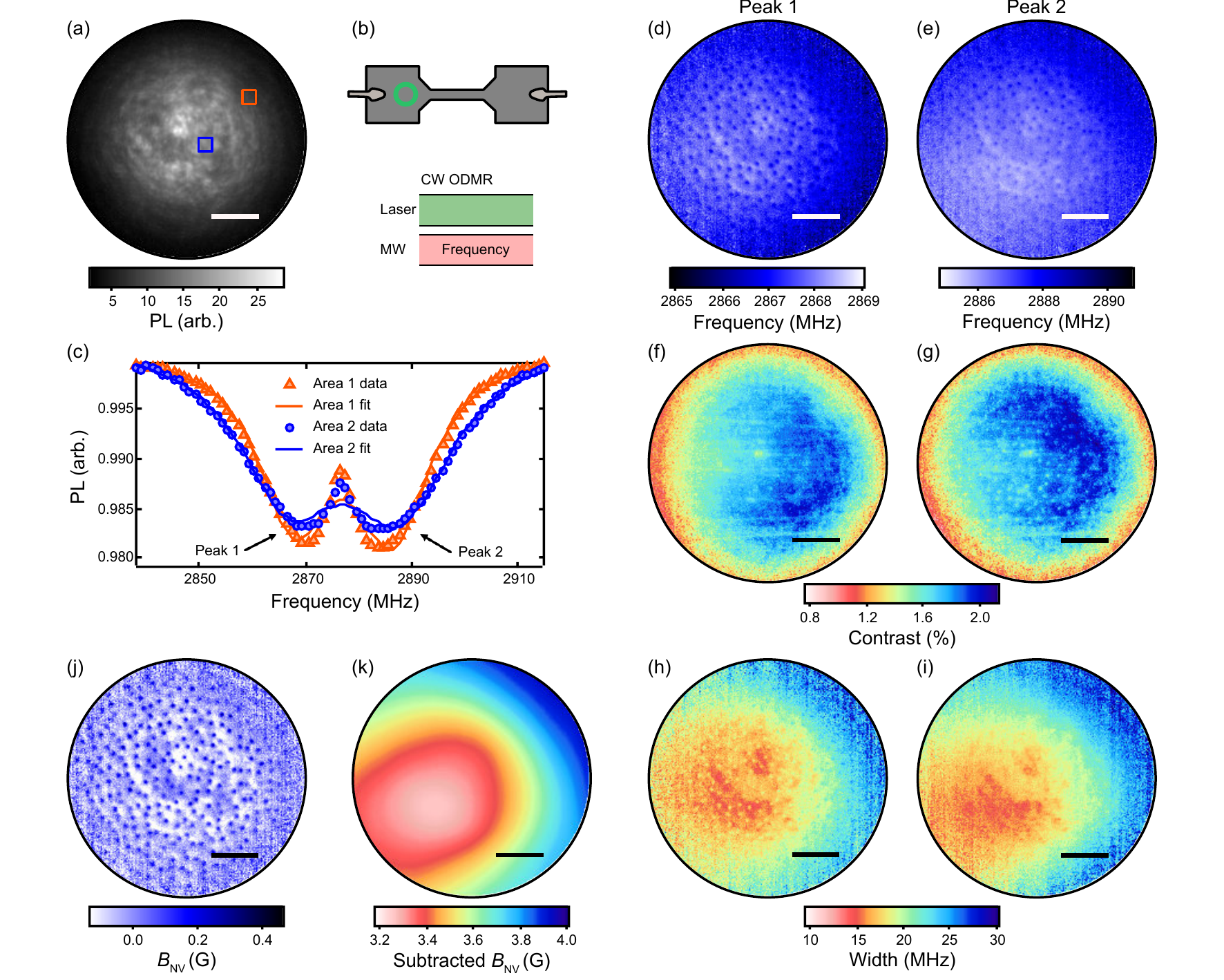}
		\caption{\textbf{Low-field ODMR imaging of vortices:} (a) PL image of the region in which vortices are imaged. This is the same region imaged in Fig.\,2 in the main text. (b, upper) Illustration showing the location of the imaged area (green circle). (b, lower) Schematic of the CW ODMR measurement sequence. (c) Low-field CW ODMR sprectra data and fits from two areas within the field of view, at higher (blue, circles) and lower local laser power (orange, triangles). Their locations within the field of view are shown in (a). (d) Frequency, (f) optical contrast, and (h) peak width of the lower frequency resonance in the low-field ODMR spectrum, fit across the same field of view as shown in (a). (g), (i), and (k) show the same fit parameters for the higher frequency peak. (j) Map of the magnetic field projected along the NV axis, $B_\text{NV}$, determined from the splitting of the two peaks identified in (c), and applying a smoothing filter. (k) The slowly varying $B_\text{NV}$ contour subtracted from the frequency splitting by the smoothing filter to produce (j). The contour is fit to Gaussian features varying over $20$\, pixels or greater on the raw $B_\text{NV}$ map. All scale bars are $20$\,$\mu$m.}
		\label{FigSIlowfield}
	\end{center}
\end{figure*}

CW ODMR images of vortices were produced by PL accumulation over a ($100$\,$\mu$m)$^2$ field of view [Fig.\,\ref{FigSIlowfield}(a)], within the large bonding pad of the Nb film [Fig.\,\ref{FigSIlowfield}(b) upper], synchronised with the continuous laser illumination and microwave (MW) field at a given MW frequency [Fig.\,\ref{FigSIlowfield}(b) lower]. ODMR spectra in this region, acquired under a field $B_0 = 1.5$\,G in the $z$-direction, exhibit two broad resonances corresponding to the two electron spin transitions of the NV centres, $f_+$ and $f_-$ [Fig.\,\ref{FigSIlowfield}(c)]. In this low field regime the exact line shape of these two resonances is non-trivial due to the ensemble nature of the measurement (ensemble of NVs with different symmetry axes and local charge/spin/strain environments) and due to fluctuations of each NV’s environment (charge and spin). In particular, the narrow feature separating the two resonances has been explained by Mittiga et al.,\cite{Mittiga2018} as arising from random distributions of both local electric field (caused by charge fluctuations in nearby defects) and local magnetic field (spin fluctuations). While in principle it is possible to use the model from Ref.\,\cite{Mittiga2018} to fit our data and extract the mean value of the magnetic field, we found the convergence of the fit to be very sensitive to the initial guess, and so were not able to obtain satisfying fits across the whole set of data ($>10,000$ spectra per image, each having slightly different microwave broadening conditions, different local environments etc.). Instead, we fitted our data with a simple sum of two Lorentzian functions with independent frequencies, amplitudes and widths (solid lines in Fig.\,\ref{FigSIlowfield}(c)). While this fit function does not capture the narrow central feature and as a result may lead to systematic errors in the estimation of the magnetic field, we found this to be the most robust and sensitive method and so is well suited to the imaging of small magnetic features such as vortices.

The fit parameters for each resonance, obtained by fitting spectra across an entire image, show long-range variations that are somewhat correlated to the laser intensity distribution [Fig.\,\ref{FigSIlowfield}(d) - (i)]. These variations are thus attributed to fitting errors due to the simplified model as explained above. For instance, the fits to the two spectra in Fig.\,\ref{FigSIlowfield}(c) indicates frequencies ($f_- = 2867.8$\,MHz, $f_+ = 2886.4$\,MHz) for Area 1, against ($f_- = 2866.8$\,MHz, $f_+ = 2888.0$\,MHz) for Area 2, but the difference can be explained by a change in the width of the resonances, caused mainly by a difference in laser intensity between the two areas, rather than by an actual difference in magnetic field. The resulting magnetic field is calculated from the fit frequencies, by the approximate formula $B_\text{NV} = (f_+-f_-)/(2\gamma_e)$,\cite{Rondin2014} [Fig.\,\ref{FigSIlowfield}(j)]. The non-physical long-range variations that carry over from the frequency maps have been removed by applying a smoothing filter to the image, which fits features in the raw $B_\text{NV}$ map that vary over more than $20$\,pixels [Fig.\,\ref{FigSIlowfield}(k)] and subtracts them, leaving only the short range features (i.e. the vortices). All the data in Fig.\,2 of the main text have been obtained by applying the fitting and post-processing method outline above. Note that the $B_\text{NV}$ thus extracted corresponds to the average projection of the magnetic field along the four different NV axes, since the different NV families are not resolved in the ODMR spectra. For a field pointing in the $z$-direction (i.e. normal to the diamond surface, which is the case at the centre of the vortices), the projection is identical for all four NV families in our $\langle100\rangle$-oriented diamond, given by $B_\text{NV} = B_z/\sqrt{3}$. The $B_z$ images in Fig.\,2 of the main text have been scaled accordingly.

\begin{figure}[ht]
	\begin{center}
		\includegraphics[width=1\columnwidth]{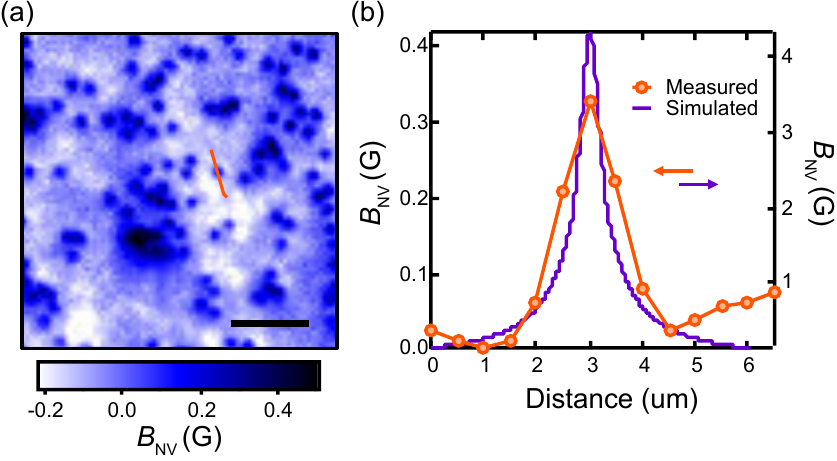}
		\caption{\textbf{Measured and simulated vortex size}: (a) $B_\text{NV}$ image of vortices, indicating the location of the line cut across a single vortex (orange). The vortices were imaged at $P_\text{laser} = 0.5$\,mW, and $B_0 = 1.5$\,G. The scale bar is $10$\,$\mu$m. (b) $B_\text{NV}$ profile across a single vortex as taken from the image in (a) (orange, left axis), and simulated (purple, right axis).}
		\label{FigSI_vortex_res}
	\end{center}
\end{figure}

The magnetic images produced by the method outlined above, allow us to determine the size and magnitude of the magnetic field features associated with the vortices. Examining our $B_\text{NV}$ images over smaller areas better highlights the features of individual vortices [Fig.\,\ref{FigSI_vortex_res}(a)]. A line profile through one such vortex shows the peak measured field projected along the NV axis is approximately $\sim0.3$\,G, with a full width at half maximum (FWHM) of $1.4$\,$\mu$m [Fig.\,\ref{FigSI_vortex_res}(b)], comfortably above our optical resolution limit, $\sim700$\,nm. To compare this result with theory, we calculated the expected vortex size following Carneiro et al.~\cite{Carneiro2000}. We consider a straight vortex in a $200$\,nm thick film with an isotropic London response, with London penetration depth, $\lambda = 400$\,nm, and superconducting coherence length, $\xi = 8$\,nm, at temperature $T/T_c = 0.75$, as inferred from Fig.\,3 in the main text. The NV response is accounted for by considering this field at a distance $130$\,nm below the film, which corresponds to the mean NV depths, as shown in Fig.\,\ref{FigSI_SRIM}. The field as seen by the NV is taken as the average projection of this field along each NV axis, which we can compare directly to the measured distribution [Fig.\,\ref{FigSI_vortex_res}(b)]. The simulated vortex field has a FWHM of $600$\,nm, less than half the width measured, and a peak field strength of $4$\,G, an order of magnitude above our measured value. This discrepancy in size is likely due to disorder within the film, which will alter the straightness of the vortex within the film, and hence its size, and explain the non-uniform vortex sizes across our field of view. The peak field discrepancy is partially explained by the observed broadening, but is likely dominated by our simplified fitting model, which neglects residual splitting of the NV spin-resonance lines, and hence reduces $B_\text{NV}$~\cite{Broadway2018b,Broadway2019}.

\begin{figure}[ht]
	\begin{center}
		\includegraphics[width=1\columnwidth]{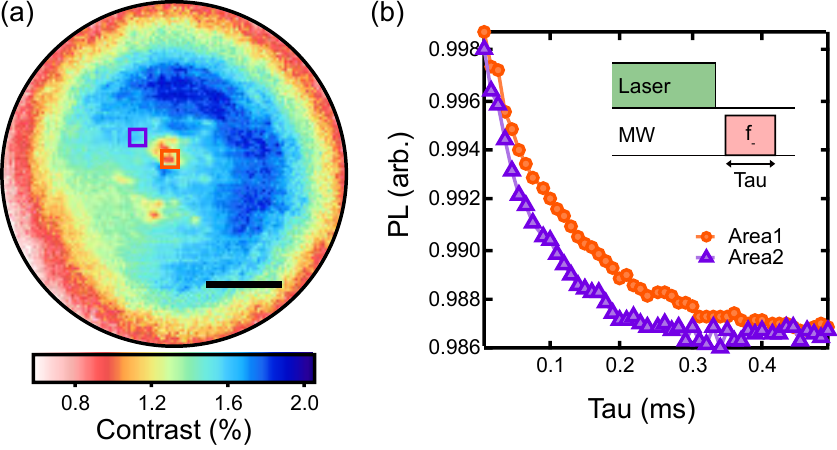}
		\caption{\textbf{Correlation between ODMR contrast and MW amplitude}: (a) PL contrast of the lower frequency resonance line ($f_-$) from a low field ODMR measurement. Two areas with are marked where the contrast is lower (orange) and higher (purple). The scale bar is $20$\,$\mu$m. (b) Rabi curves integrated over the two areas marked in (a). The Rabi oscillation frequency is greater for the area with higher ODMR PL contrast (purple), than the one with lower ODMR contrast (orange). The Rabi measurement pulse sequence uses a long laser pulse for initilisation and readout of the NV spin-state, and resonant microwave pulse or varying duration (inset). $P_\text{laser} = 1.0$\,mW and $P_\text{MW} = 2.0$\,mW for both measurements.}
		\label{FigSI_Rabi}
	\end{center}
\end{figure}

Local reductions in ODMR contrast, as seen in Fig.\,2 of the main text, are attributed to a reduction in the local microwave field strength due to the normal state Nb or superconducting Nb with a superconducting gap close to the microwave frequency. However, PL spin contrast can be reduced by a number of effects including the intensity of the excitation laser, large magnetic field gradients,\cite{Tetienne2018b} and interactions of NVs with nearby defects\cite{Bluvstein2019} and materials.\cite{Lillie2018} To validate our interpretation, we compare ODMR contrast and Rabi curve measurements of the same region, imaged at base temperature with $P_\text{laser} = 1.0$\,mW. An ODMR contrast map of an area within the Nb bonding pad taken under these conditions shows localised pockets of reduced PL contrast towards the centre of the image [Fig.\,\ref{FigSI_Rabi}(a)], similar to those seen in Fig.\,2(f). Rabi curves from these pockets show slower Rabi oscillations as compared to neighbouring regions with greater ODMR contrast [Fig.\,\ref{FigSI_Rabi}(b)], i.e. for a given interaction time with the microwave field, the spin-state dependent PL evolves less rapidly in these regions, giving up to $0.2\%$ more PL for an interaction time of $0.15$\,ms. Quantitative comparisons with ODMR contrast measurements are non-trivial given the pulsed nature of the Rabi measurement versus the CW ODMR measurement, however, this result demonstrates that locally reduced microwave field power at least partially explains the reductions in ODMR contrast. We note that the laser heating in each measurement should be comparable given that a $1$\,s initialising laser pulse was used prior to camera exposure ($0.5$\,s), which has a minimum laser duty cycle $0.2$ for the microwave pulse durations explored.

The number of vortices within a given area is such that the sum of their fluxes is equal to the total flux passing through the same area in the normal state. Each vortex is threaded by a magnetic flux quantum, $\Phi_0$, and hence $N\,\Phi_0 = \bf{B}_0 \cdot \bf{A}$, where $N$ is the number of vortices within an area, $\bf{A}$, at a magnetic field, $\bf{B}_0$. The vortex images presented in Fig.\,2 in the main text were taken at a magnetic field $B_0$ $\approx 1.5$\,G in the $z$-direction. This field is actually due to some background field and residual magnetisation within the unshielded cryostat, as there was no field applied by the superconducting coils for these measurements, i.e. $B_\text{app} = 0$\,G. Here we demonstrate the effect of varying our applied magnetic field strength, $B_\text{app}$, which differs slightly to the field seen at the sample $B_0$, given the un-shielded cryostat and remanent magnetisation.

\begin{figure}[t]
	\begin{center}
		\includegraphics[width=1.0\columnwidth]{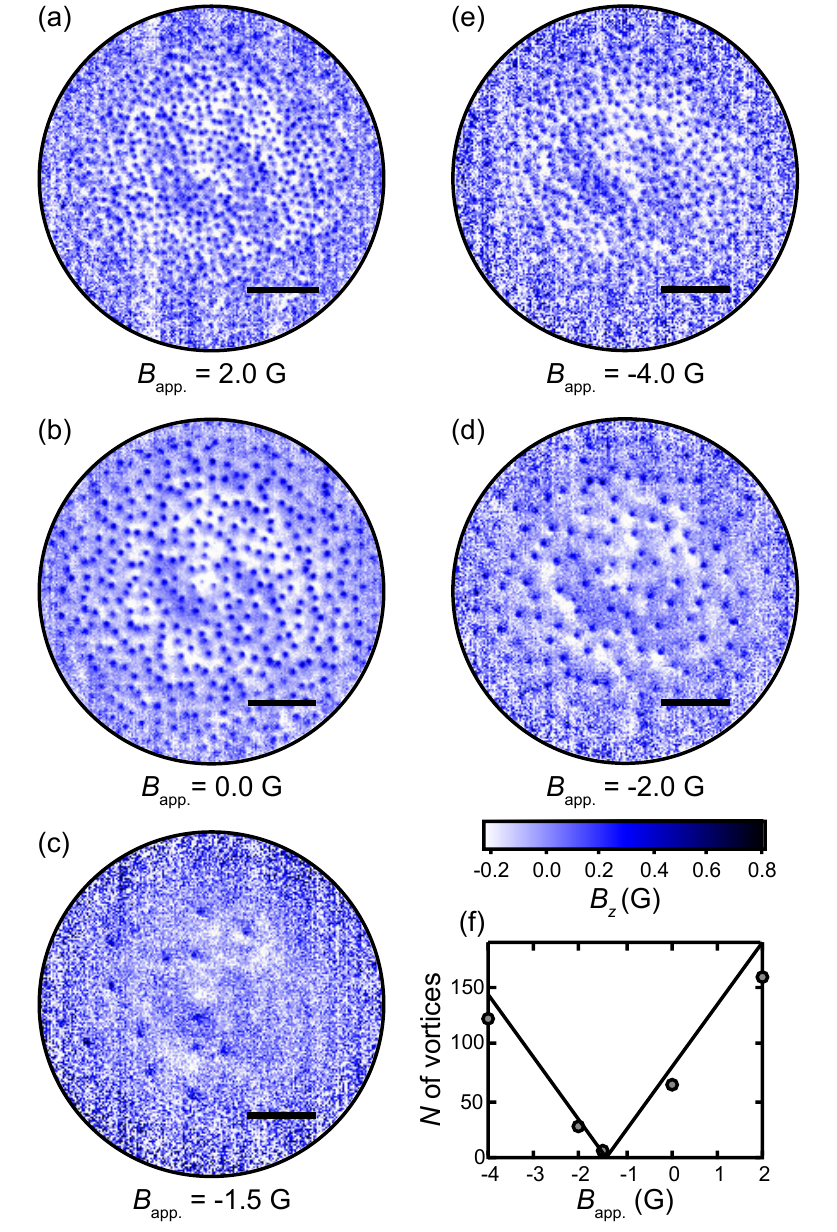}
		\caption{\textbf{Vortex imaging varying the applied field:} $B_z$ maps imaged at base temperature, with $P_\text{laser} = 0.5$\,mW, and $\bf{B}_\text{app}$ in the $z$-direction with magnitudes: (a) $2.0$\,G, (b) $0.0$\,G, (c) $-1.5$\,G. (d) $-2.0$\,G, and (e) $-4.0$\,G. The system was degaussed prior to this imaging series to minimise contributions from magnetic components within the cryostat. All scale bars are $20$\,$\mu$m. (f) Number of vortices, $N$, within a ($33.81$\,$\mu$m)$^2$ area, counted from the previous images, as a function of $B_\text{app}$. The solid line shows $N\,\Phi_0 = (B_\text{app}+1.4054\,\text{G}) \times A$.}
		\label{FigSI_field_nuc}
	\end{center}
\end{figure}

Varying $B_\text{app}$ from $2.0$ to $-4.0$\,G in the $z$-direction, the number of vortices within the field of view changes considerably [Fig.\,\ref{FigSI_field_nuc}(a)-(e)]. Counting the number of vortices within a fixed area, ($33.81$\,$\mu$m)$^2$, and fitting the number as a function of the applied field, shows a strong linear relation with an effective zero field at $B_\text{app} = -1.5(1)$\,G. Accounting for this offset in field strength, $B_0 = B_\text{app}+1.5$\,G, the number of vortices compares well with theory [Fig.\,\ref{FigSI_field_nuc}(f)], with a slight discrepancy likely due an uncalibrated magnification from a non-ideal optics setup. We note that the $B_0$ values quoted in the main text for current imaging do not include this correction to the field at the sample, as it is small compared to the applied field.

\begin{figure}[t]
	\begin{center}
		\includegraphics[width=1.0\columnwidth]{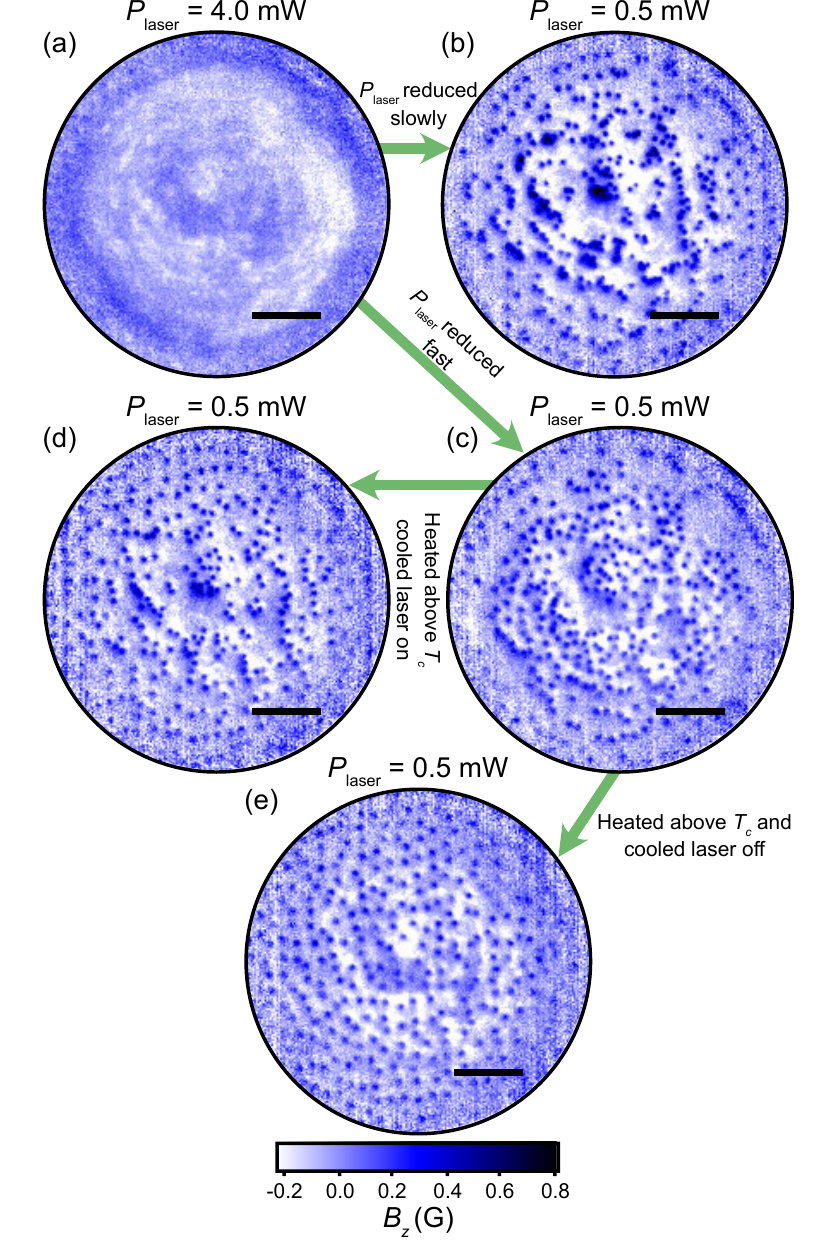}
		\caption{\textbf{Vortex clustering following various cooling procedures:} $B_z$ images of vortices at $B_0 = 1.5$\,G imaged at: (a) $P_\text{laser} = 4.0$\,mW giving a large normal state region at the centre of the image (same as Fig.\,2(d)); (b) $P_\text{laser} = 0.5$\,mW after reducing $P_\text{laser}$ from $4.0$\,mW over $\sim 5$\,seconds (same as Fig.\,2(l)); (c) $P_\text{laser} = 0.5$\,mW after cutting the $P_\text{laser} = 4.0$\,mW beam, letting $T_\text{sensor}$ stabilise at base temperature, and then imaging. (d) and (e) are imaged at $P_\text{laser} = 0.5$\,mW after heating the cryostat such that $T_\text{sensor} > T_c$ and cooling to base temperature in the presence and absence of laser respectively, at the same power respectively. All scale bars are $20$\,$\mu$m.}
		\label{FigSI_laser_cool}
	\end{center}
\end{figure}

Vortex images shown in the main text demonstrate clustering of vortices around local hot-spots when the laser power is reduced from powers which quench superconductivity, $P_\text{laser} \geq 1.0$\,mW, to a less invasive imaging power, $P_\text{laser} = 0.5$\,mW. The final arrangement of vortices after cooling is sensitive to both the initial laser power, as shown in the main text in Figs.\,2(k) and (l), but also the rate at which the laser power is reduced. Reducing the laser power from $P_\text{laser} = 4.0$\,mW [Fig.\,\ref{FigSI_laser_cool}(a)] to $P_\text{laser} = 0.5$\,mW by turning down the power over a few seconds and then imaging at $P_\text{laser} = 0.5$\,mW [Fig.\,\ref{FigSI_laser_cool}(b)] results in greater clustering than if the beam is cut immediately (in $\approx10$~ns using the acousto-optic modulator) and then imaged at $P_\text{laser} = 0.5$\,mW once $T_\text{sensor}$ has stabilised [Fig.\,\ref{FigSI_laser_cool}(c)]. Similar clustering is observed when the cryostat is cooled to base temperature from $T_\text{sensor} > T_c$ with the laser on, at $P_\text{laser} = 0.5$\,mW [Fig.\,\ref{FigSI_laser_cool}(d)]. A uniform vortex configuration is recovered when the cryostat is heated above $T_c$ and cooled to base temperature in the absence of laser [Fig.\,\ref{FigSI_laser_cool}(e)], indicating that the vortex clustering is rewritable.

\section{Theoretical modelling}

In the following, we describe how to model superconducting vortices in a temperature gradient, induced for instance by heating a sample locally with a laser.
 
\begin{figure}[htb]
 \includegraphics[width=0.9\columnwidth]{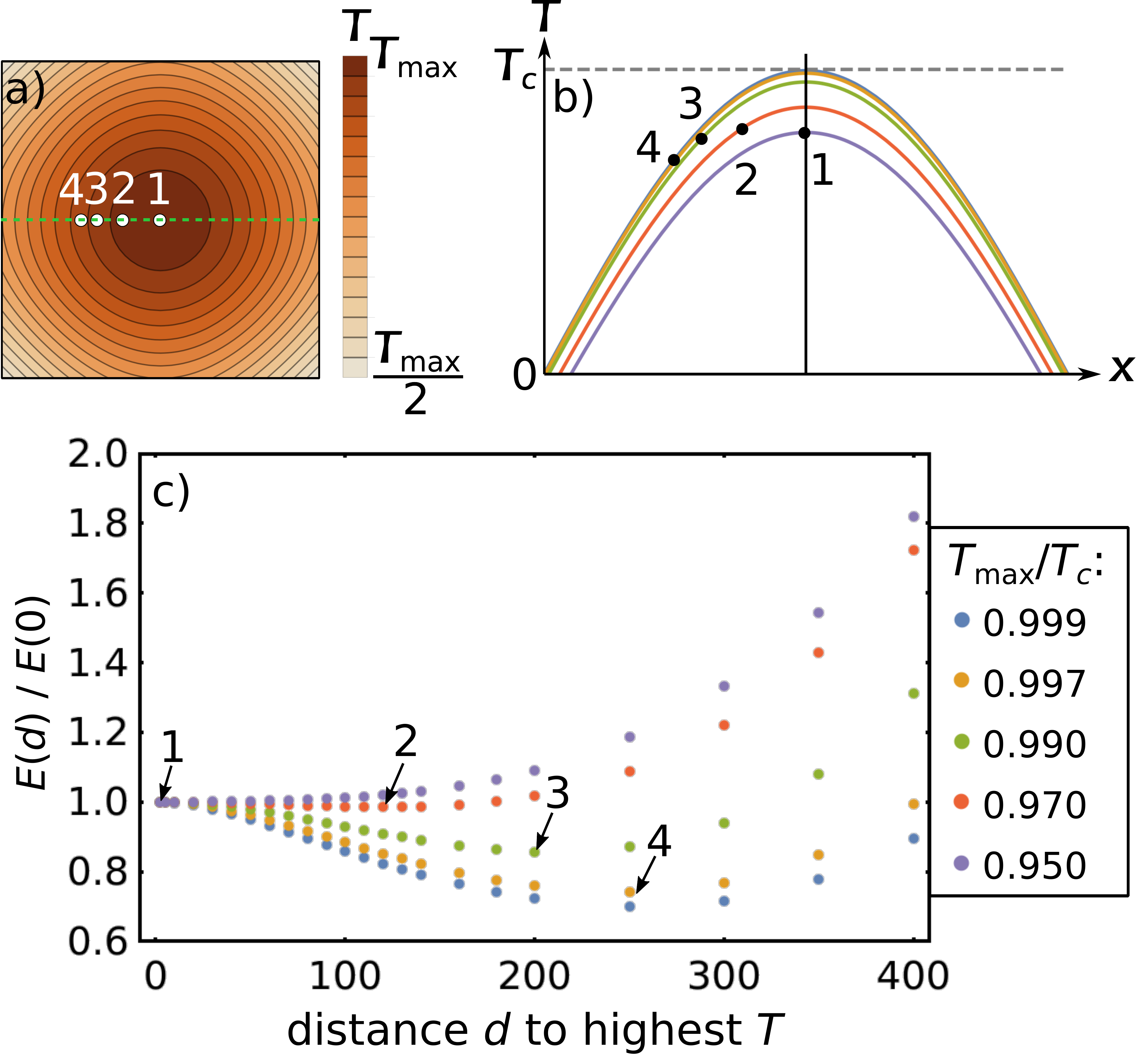}
 \caption{\label{fig:Vort_in_gradT} Simulations of a single vortex in a temperature gradient according to Eq.\,\eqref{eq:Tgrad}, where we set $\lambda(0)=20$. The numbers 1 to 4 indicate the positions of the single vortex relative to the point with the highest temperature that yield minimal energy. Number 4 denotes the distance for both $T_{\rm max}=0.999\,T_{c}$ and $T_{\rm max}=0.997\,T_{c}$. (a) Contour plot of the temperature gradient. (b) Linear cut through the temperature profile as indicated by the green dashed line in (a) for different values of the highest temperature, $T_{\rm max}/T_{c}$. Different colors correspond to the values given in the legend of (c). (c) Energy of the single vortex as a function of the distance $d$ of the vortex from the point with the highest temperature, relative to the energy at zero distance.}
\end{figure}

 We consider a rectangular geometry with periodic boundary conditions, where we define a periodic temperature gradient (shown in Fig.\,\ref{fig:Vort_in_gradT}(a,b)) given by
 \begin{align}
 \label{eq:Tgrad}
  \tau(x,y)&=a_0+\bigg[\cos\Big(x\frac{2\pi}{L_x}\Big)+\cos\Big(y\frac{2\pi}{L_y}\Big)-2\bigg]\frac{a_0-a_1}{4},\\
  \tau(x,y)&=1-\frac{T(x,y)}{T_{c}},
 \end{align}
where $L_{x}$ and $L_{y}$ denote the length of the system in $x$- and $y$-direction, respectively. $a_{0}$ ($a_{1}$) denotes the highest (lowest) appearing value of $\tau$. Eq.\,\eqref{eq:Tgrad} is constructed such that $\tau=a_{0}$ in the corners of the rectangle and $\tau=a_{1}$ in its middle, i.e.
\begin{align*}
 \tau(0,0) = \tau(0,L_y) = \tau(L_x,0) = \tau(L_x,L_y) &= a_0,\\[6pt]
 \tau(L_x/2,L_y/2) &= a_1.
\end{align*}
Using the London equation\cite{London1935},
\begin{equation}
\label{eq:London_1}
\vec{h}(x,y)+\lambda^{2}\nabla\times\nabla\times\vec{h}(x,y)=0,
\end{equation}
we can calculate the magnetic field, $\vec{h}$, and ultimately the energy of a given configuration using the London energy functional which is given by
\begin{equation}
\label{eq:E_from_h}
 E=\int\big[\vec{h}^{2}+(\lambda\nabla\times\vec{h})^{2}\big]{\rm d}\vec{A}.
\end{equation}
With the Maxwell equation $\nabla\cdot\vec{h}=0$, the temperature dependance of the penetration length, $\lambda=\lambda(0)/\sqrt{\tau}$, and $\vec{h}=h\hat{e}_{z}$, Eq.\,\eqref{eq:London_1} becomes
\begin{equation}
\label{eq:diff_eq_hom}
 \tau h-\lambda(0)^{2}\nabla^{2}h=0.
\end{equation}

Implementing an arbitrary magnetic flux configuration $\Phi(x,y)$ into the system is simply done by adding it as a inhomogeneity to the right hand side of Eq.~\eqref{eq:diff_eq_hom}, which leads to
\begin{equation}
 \tau h-\lambda(0)^{2}\nabla^{2}h=\Phi(x,y).
\end{equation}
The integral of $\Phi$ over the whole space yields the total flux introduced. For a single vortex at position $\vec{r}_{0}$ we add the flux in form of a delta distribution, $\Phi(\vec{r}_{0})=\Phi_{0}\delta(\vec{r}_{0}-\vec{r})$, with the flux quantum $\Phi_{0}$. Multiple vortices at positions $\vec{r}_{i}$ are introduced by simply summing up their respective inhomogeneities, i.e.
\begin{equation}
 \Phi(x,y) = \Phi_{0}\sum_{i}\delta(\vec{r}_{i}-\vec{r}).
\end{equation}
Since we calculate the energy of a given vortex configuration from the total magnetic field, using Eq.\,\eqref{eq:E_from_h}, vortex-vortex interactions are automatically accounted for. This energy calculation can then be used for a Monte Carlo simulation to find the ground state for an arbitrary temperature gradient and for any number of vortices.

\subsection*{Results}

The results of the simulations for the total energy with a single vortex are shown in Fig.\,\ref{fig:Vort_in_gradT}(c). We see that as long as the temperature is sufficiently below the critical temperature, $T_{c}$, the vortex will always move towards the region with the highest temperature. When the temperature gets close to $T_{c}$ in some region, however, the vortex will be repelled from this region at short distances on the length scale of roughly 10\,$\lambda(0)$.

\begin{figure}[t]
 \includegraphics[width=0.95\columnwidth]{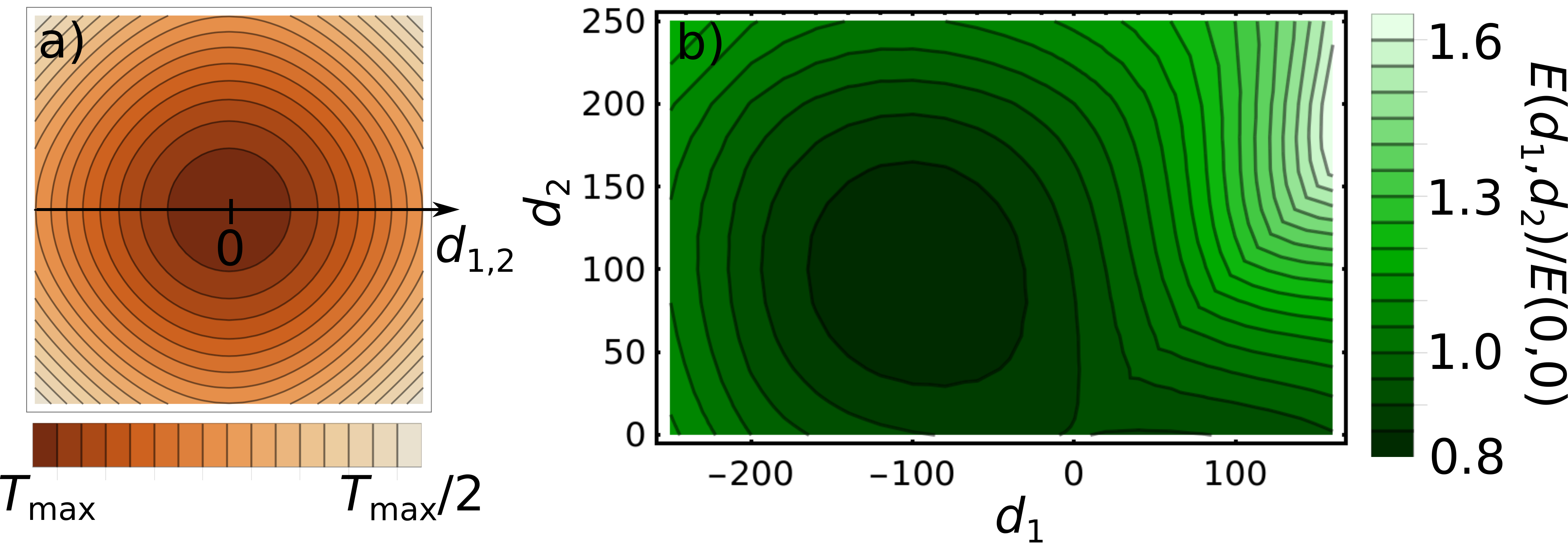}
 \caption{\label{fig:2Vort_in_gradT} Simulations of two vortices in a temperature gradient according to Eq.\,\eqref{eq:Tgrad}, where we set $\lambda(0)=20$. a) Contour plot of the temperature gradient. b) Energy of the two vortex state as a function of their distances, $d_{1}$ and $d_{2}$, from the point with the highest temperature, relative to the energy at $d_{1}=d_{2}=0$.}
\end{figure}

For a two vortex state the minimum in energy according to Eq.\,\eqref{eq:E_from_h} is given when the vortices are about 10\,$\lambda(0)$ apart from each other with the hot spot in the middle, which is shown in Fig.\,\ref{fig:2Vort_in_gradT}. Furthermore, the shown energy landscape implies that, if both vortices were on the same side of the hot spot, the vortex closer to the hot spot would get pushed through to the other side of it.

The behavior of a system with many vortices can be extrapolated from the results shown here. Vortices still tend to move towards regions with higher temperature, but the repulsive vortex-vortex interaction keeps them apart and thus acts as a limiting force. This will result in clustering around hot spots for a low number of vortices and a gradient in the vortex density for a very large number of vortices, where the density is higher in warmer regions.

When the sample is heated locally (assuming adiabatically slow temperature change), as done in the experiments with a laser, the vortex density would first increase in the hot region. When the temperature becomes close to $T_{c}$, however, the vortices not only get repelled from each other but also from the center of the hot spot (based on Fig.\,\ref{fig:Vort_in_gradT}). We can safely assume that with an increasing vortex density close to the hot spot, eventually it becomes energetically favourable for the many-vortex system to push some of the vortices into the hot spot and, when the temperature at the hot spot exceeds $T_{c}$, into the normal conducting region. 

Likewise, when the temperature decreases and the normal conducting region shrinks, it will become energetically favourable for the vortices to re-nucleate into the superconducting region where they will remain trapped near the energy minimum seen in Fig.\,\ref{fig:Vort_in_gradT}(c). Once the temperature becomes sufficiently far below $T_c$ everywhere\cite{Stan2004}, pinning becomes the dominant effect and so the vortex clusters become frozen around the former hot spots. This is the essence of the mechanism that leads to the clustering observed in our experiments. We note that pinning centers can be simply included in the above model by an additive pinning potential on top of the energy function shown in Figs.\,\ref{fig:Vort_in_gradT} and and \ref{fig:2Vort_in_gradT}.

\section{Imaging of transport currents}\label{vectorODMR}

\begin{figure*}[ht]
	\begin{center}
		\includegraphics[width=0.8\textwidth]{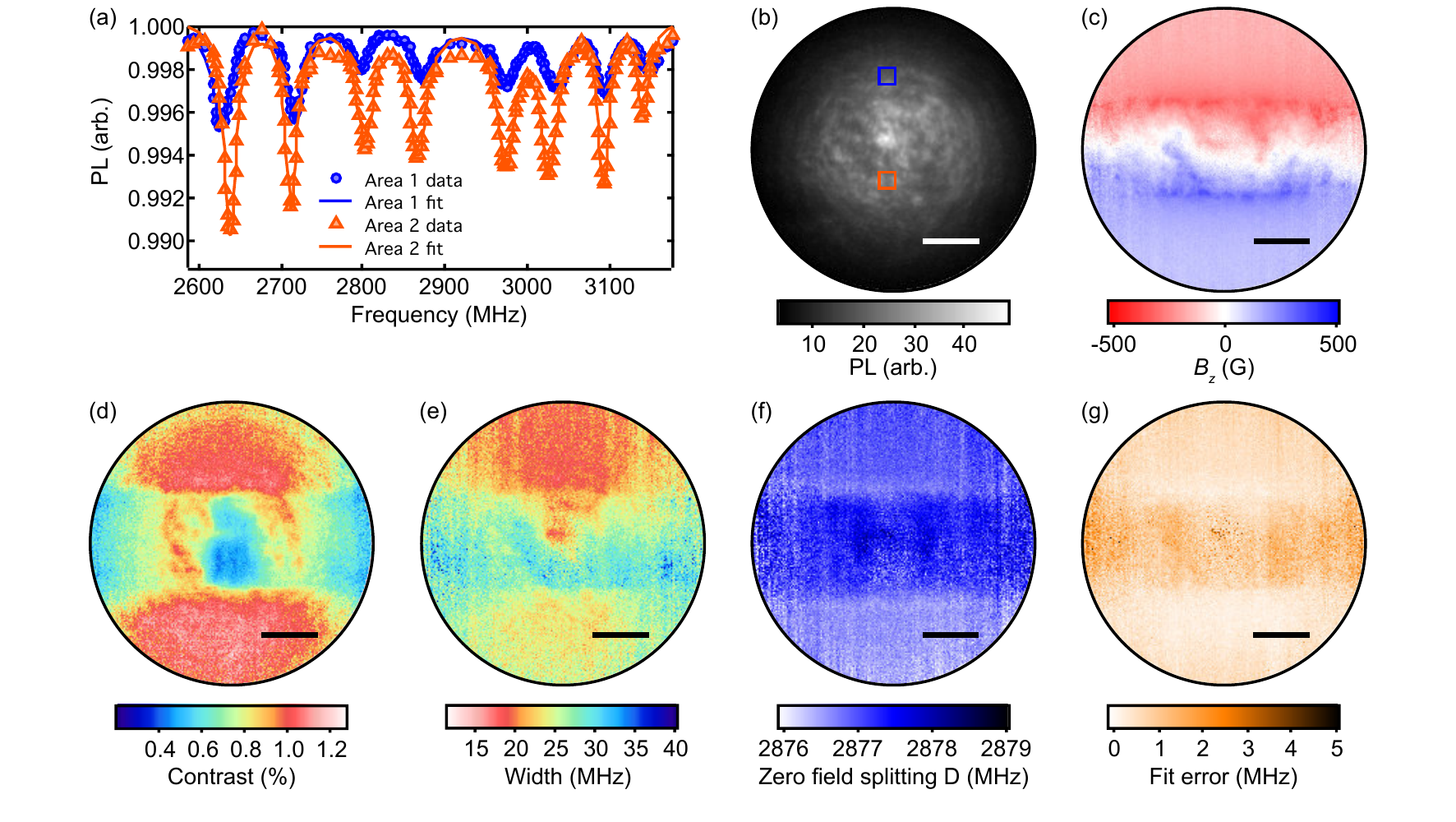}
		\caption{\textbf{Vector magnetometry for imaging transport currents:} (a) ODMR sprectra from two areas, one beside the Nb wire (blue), and one under the wire close the the edge (orange). (b) PL image of the region measured, which is the same as in Fig.\,4 of the main text. The image is taken under CW illumination at $P_\text{laser} = 0.5$\,mW. The location of the areas for which the spectra are given in (a) are shown. (c) $B_z$ map inferred from the analysis described in the text. (d) and (e) show the contrast and width, respectively, for the lowest frequency transition of the ODMR spectrum. The equivalent maps for the other seven resonance lines are qualitatively similar. (f) Map of the zero-field splitting parameter, $D$. (g) Residual error of the fit, $\varepsilon$. The scale bars are $20\,\mu$m.}
		\label{FigSIvectorODMR}
	\end{center}
\end{figure*}

In order to image transport currents, as shown in Fig.\,4 of the main text, we need to accurately measure the magnetic field projection (including the sign) of at least one NV family out of the four available. This requires to apply a bias magnetic field of at least $40$\,G typically (for the diamond sample used here) to separate one NV family from the other three in the ODMR spectrum. Moreover, to improve the accuracy of the reconstruction it is preferable to measure all four NV families simultaneously rather than just one.\cite{Tetienne2017,Tetienne2019} This can be achieved by aligning the bias magnetic field such that all four NV families can be resolved in the ODMR spectrum. Here we apply a magnetic field $\bf{B}_\text{app}=\rm{(47.5,97.4,19.1)}$\,G expressed in lab frame coordinates (see definition of the $xyz$ axes in Fig.\,1(e) of the main text). In this frame the NV axes have unit vectors $\bf{u}_\text{NV}=\rm{(\pm1,\pm1,1)/\sqrt{3}}$. Example ODMR spectra recorded under this field are shown in Fig.\,\ref{FigSIvectorODMR}(a).

To analyse the ODMR data, we first fit the spectrum at each pixel with a sum of eight Lorentzian functions with free frequencies, amplitudes and widths (solid lines in Fig.\,\ref{FigSIvectorODMR}(a)). The eight resulting frequencies $\{f_i\}_{i=1\dots8}$ are then used to infer the total magnetic field, $\bf{B}_{\rm tot}$, by minimising the root-mean-square error function
\begin{equation} \label{Eq:error}
\varepsilon(D,{\bf{B}}_\text{tot}) = \sqrt{\frac{1}{8}\sum_{i=1}^8 \left[f_i - f_i^{\rm calc}(D,\bf{B}\rm_{tot}) \right]^2}
\end{equation}
where $\{f_i^{\rm calc}(D,\bf{B}\rm_{tot})\}_{i=1\dots8}$ are the calculated frequencies obtained by numerically computing the eigenvalues of the spin Hamiltonian for each NV orientation,
\begin{eqnarray} \label{eq:Ham}
{\cal H} &= DS_Z^2+\gamma_{\rm NV}\bf{S}\cdot\bf{B}~,
\end{eqnarray}
and deducing the electron spin transition frequencies. Here $\bold{S}=(S_X,S_Y,S_Z)$ are the spin-1 operators, $D$ is the temperature-dependent zero-field splitting, $\gamma_\text{NV}=28.035(3)$\,GHz/T is the isotropic gyromagnetic ratio, and $XYZ$ is the reference frame specific to each NV orientation, $Z$ being the symmetry axis of the defect~\cite{Doherty2012,Doherty2013} with unit vector $\bf{u}_{\rm NV}$ defined previously. The resulting total magnetic field $\bf{B}_{\rm tot}$ contains both the applied bias field and the field from the Nb film due to the magnetic response as well as the transport current (\O rsted field). Since the applied field is uniform over the field of view, we simply subtract a constant offset to $\bf{B}_{\rm tot}$, taken to be the field measured far from the Nb wire, to obtain the Nb-induced field. Illustrative results of this analysis are shown in Fig.\,\ref{FigSIvectorODMR}, corresponding to the case studied in Fig.\,4(b) and (e) of the main text ($I=20$\,mA, $P_{\rm laser}=0.5$\,mW): the PL image is shown in (b), the background-subtracted $B_z$ map in (c), the contrast and width of the lowest ODMR resonance in (d) and (e), the $D$ parameter in (f) and the residue $\varepsilon$ in (g). The fit residue seems to correlate with the contrast: $\varepsilon<1$\,MHz beside the Nb wire but it reaches $2$\,MHz under the Nb where the ODMR contrast is lower as can be seen in the ODMR spectra of Fig.\,\ref{FigSIvectorODMR}(a). This suggests that $\varepsilon$ is dominated by the noise in the data and that the model Eq.~(\ref{eq:Ham}) is adequate. In comparison, the shifts induced by the Oersted field are much larger, up to $10$\,MHz. The zero-field splitting parameter $D\approx2877$\,MHz is roughly as expected at this temperature,\cite{Doherty2014a} with a step of $\sim0.5$\,MHz under the Nb wire attributed to strain.\cite{Broadway2019} 

\begin{figure}
	\begin{center}
		\includegraphics[width=1.0\columnwidth]{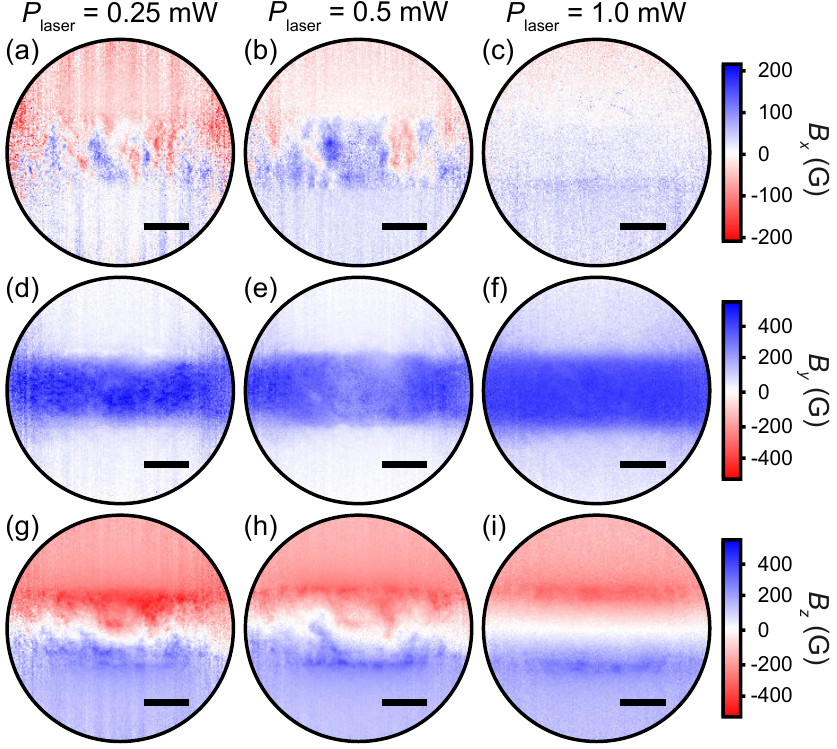}
		\caption{\textbf{Vector magnetic field:} Maps of the background-subtracted vector magnetic field components $B_x$ (a,b,c), $B_y$ (d,e,f), and $B_z$ (g,h,i) for the three cases studied in Fig.\,4 of the main text: $P_\text{laser}=0.25$\,mW (left column), $P_\text{laser}=0.5$\,mW (middle column), $P_\text{laser}=1.0$\,mW (right column). The scale bars are $20\,\mu$m.}
		\label{FigSIvectormag}
	\end{center}
\end{figure}

\begin{figure}
	\begin{center}
		\includegraphics[width=1.0\columnwidth]{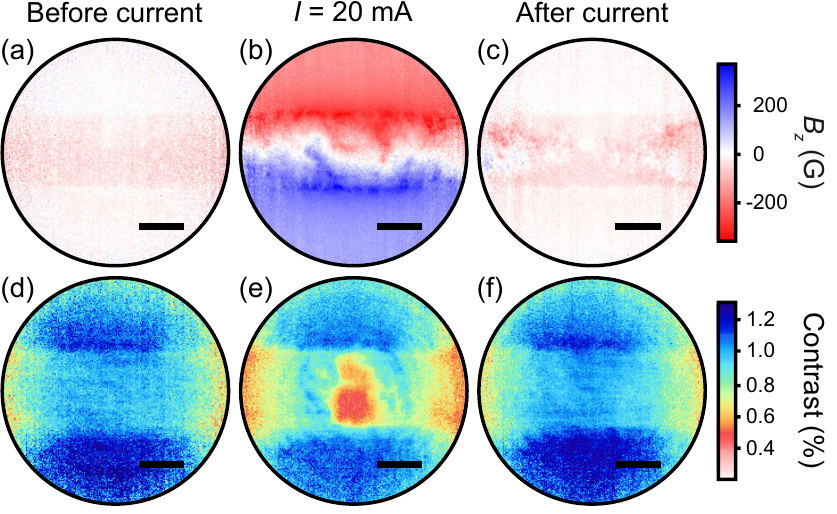}
		\caption{\textbf{Magnetic response of the Nb wire:} Maps of the background-subtracted $B_z$ magnetic field component (top row) and ODMR contrast (bottom row) measured with $P_{\rm laser}=0.5$\,mW and ${\bf B}_{\rm app}=(47.5,97.4,19.1)$\,G. (a) was taken without current just after cooling down, (b) corresponds to the case $I=20$\,mA shown before, (c) was taken just after turning off the current. The scale bars are $20\,\mu$m.}
		\label{FigSInocurrent}
	\end{center}
\end{figure}

The three components $(B_x,B_y,B_z)$ of the magnetic field are shown in Fig.\,\ref{FigSIvectormag} for the three cases studied in Fig.\,4 of the main text. To separate the \O rsted field from the magnetic response of the Nb film, we performed the same measurements prior to applying the transport current (just after cooling the sample below $T_c$) and after turning off the current following the measurement with current presented before. The $B_z$ maps for those three cases are shown in Fig.\,\ref{FigSInocurrent}, along with corresponding maps of the ODMR contrast. Upon cooling in the applied field of $\bf{B}\rm_{app}=(47.5,97.4,19.1)$\,G, the field exhibits a small expulsion as apparent from a reduction of $B_z$ by $\approx0.4$\,G under the Nb wire (Fig.\,\ref{FigSInocurrent}(a)). The expected density of vortices due the applied $B_z$ is about $1$ vortex$/\mu$m$^2$,\cite{Stan2004} hence the weak expulsion is in broad agreement with the London penetration depth estimated from the device resistance ($\lambda\approx 400$\,nm). Upon turning the current on, the field becomes dominated by the \O rsted contribution, reaching $B_z\pm5$\,G (Fig.\,\ref{FigSInocurrent}(b)). Turning off the current, the small field expulsion is recovered but some additional features have appeared, indicating a rearrangement of the vortices during the application of the current. Because this response field may be different with the current on, and because in any case it remains much smaller than the \O rsted field, we will simply ignore it when reconstructing the current density. That is, the reconstructed current density will also include a small contribution from the supercurrent associated with the response to the applied field. This supercurrent is localised near the edges of the wire and has opposite signs for each edge, so that the net current integrated across the width of the wire will be only due to the transport current. 

Comparing the ODMR contrast maps in Fig.\,\ref{FigSInocurrent}, a drop in contrast can be seen at the centre of the wire with the current on, whereas the contrast is relatively uniform in the absence of current. From the observations made in Fig.\,2 of the main text, we tentatively ascribe this drop in contrast to a portion of the Nb wire turning normal, but not extending through the entire thickness of the film since the resistance remains $R=R_\text{SC}$. In the vortex imaging presented in Fig.\,2 of the main text, there was no significant change in local contrast at $P_\text{laser}=0.5$\,mW, however here the presence of a relatively large current density may lower $T_c$ sufficiently to allow the local heating due to the laser to locally quench the superconductivity. 

We now describe the methods for reconstructing the current density in the Nb film. In principle, the three magnetic field components $(B_x,B_y,B_z)$ are related to each other via Amp{\`e}re's law, $\nabla\cross{\bf B}=0$,\cite{Lima2009} and there are several ways (nominally equivalent) to use them to reconstruct the current density. For instance, the projected current density $\tilde{\bf J}=\int \bf J \,\rm dz$ can be obtained from the in-plane components $B_x$ and $B_y$ using~\cite{Tetienne2019} 
\begin{eqnarray}
\tilde{J}_x &=& -\frac{2}{\mu_0}B_y \label{eq:Jx} \\
\tilde{J}_y &=& \frac{2}{\mu_0}B_x, \label{eq:Jy}
\end{eqnarray}
where we utilised the fact that the Nb-NV distance (between $0$\,-\,$400$\,nm) is small compared to the lateral spatial resolution of the measurements ($\approx1\,\mu$m given by the optical resolution).\cite{Tetienne2019} Below, this method will be referred to as the first method. Alternatively, we can use the $B_z$ component together with the continuity of current ($\nabla\cdot{\bf J}=0$) to obtain relations in the Fourier plane,
\begin{eqnarray}
\tilde{\mathcal{J}}_x &=& -\frac{2}{\mu_0}\frac{ik_y}{k}\mathcal{B}_z \label{eq:jx} \\
\tilde{\mathcal{J}}_y &=& \frac{2}{\mu_0}\frac{ik_x}{k}\mathcal{B}_z. \label{eq:jy}
\end{eqnarray}
Here $\mathcal{F}(k_x,k_y)$ denotes the two-dimensional Fourier transform of $F(x,y)$, where ${\bf k}=(k_x,k_y)$ is the spatial frequency vector and $k=\sqrt{k_x^2+k_y^2}$. This is the second method.

\begin{figure}
	\begin{center}
		\includegraphics[width=1.0\columnwidth]{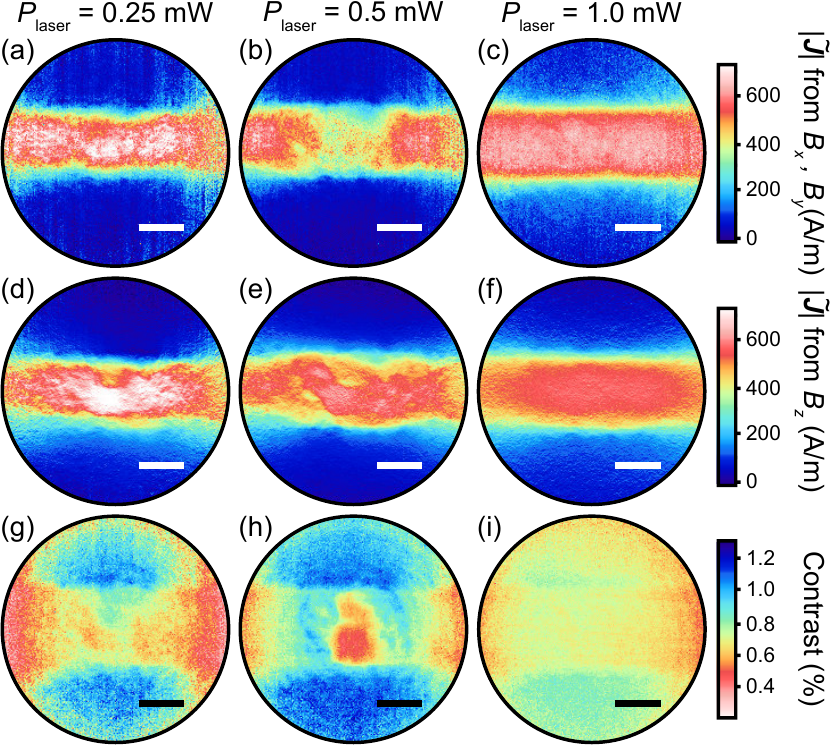}
		\caption{\textbf{Reconstructed current density:} Norm of the projected current density, $|\tilde{\bf{J}}|$, reconstructed from Eq.~(\ref{eq:Jx},\ref{eq:Jy}) (a,b,c) and from Eq.~(\ref{eq:jx},\ref{eq:jy}) (d,e,f), and corresponding ODMR contrast maps (g,h,i), for the three cases studied in Fig.\,4 of the main text: $P_\text{laser}=0.25$\,mW (left column), $P_\text{laser}=0.5$\,mW (middle column), and $P_\text{laser}=1.0$\,mW (right column). All scale bars are $20\,\mu$m.}
		\label{FigSIreconstruction}
	\end{center}
\end{figure} 

\begin{figure}[h]
	\begin{center}
		\includegraphics[width=1.0\columnwidth]{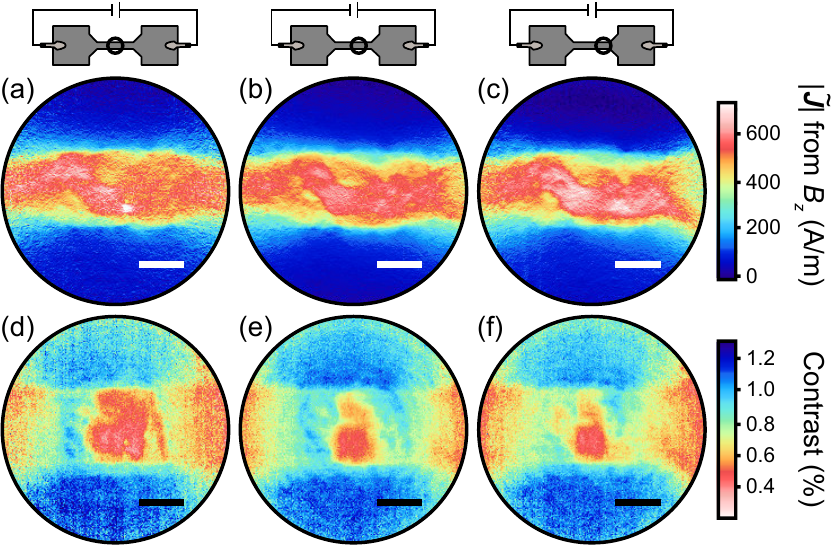}
		\caption{\textbf{Current density vs laser location:} (a,b,c) Maps of the current density, $|\tilde{\bf{J}}|$, and (d,e,f) corresponding ODMR contrast, for three different locations along the Nb strip, indicated on the diagram (above). The same current, $I = 20$\,mA, and laser power, $P_\text{laser}=0.5$\,mW, were used for each image. The scale bar is $20\,\mu$m.}
		\label{FigSIJvsLocation}
	\end{center}
\end{figure}

The projected current density reconstructed from these two methods are shown in Fig.\,\ref{FigSIreconstruction} for the three cases presented in Fig.\,4 of the main text. We note that the current densities shown in the main text are in fact projected current densities, labeled as $\bf{J}$ (without the tilde) for simplicity. There are several differences between the two methods. Considering the case where the Nb wire is in the normal state (Fig.\,\ref{FigSIreconstruction}(right column)), we see that the first method gives a $|\tilde{\bf{J}}|$ map that looks sharper and more uniform along the Nb wire than the second method, for which the current appears tapered near the left and right boundaries of the image. This is because the latter is prone to truncation artefacts when calculating the Fourier transform of $B_z$, and those are more pronounced near the boundaries where the $B_z$ data has more noise. In contrast, there is no Fourier transform involved in the first method and so no truncation artefact; the noise in $|\tilde{\bf{J}}|$ directly mirrors the noise in the $B_x$, $B_y$ data. While this observation would normally motivate the use of the first method, examination of the maps in Fig.\,\ref{FigSIreconstruction}(middle column) reveals another effect. With the first method, there appears to be a ``missing'' current near the centre of the image. Indeed, integrating the $x$-component of the current density across the width of the wire, i.e. $I_{\rm int.}=\int \tilde{J}_x\,dy$, we approximately recover the electrically measured current of $I=20$\,mA near the boundaries of the image but the value of $I_{\rm int.}$ drops to $14$\,mA (a $30$\% drop) near the centre. This corresponds to a drop in $B_y$ seen in Fig.\,\ref{FigSIvectormag}(b) and also correlates with a drop in ODMR contrast observed in this region (third column in Fig.\,\ref{FigSIreconstruction}) indicating a local quenching of superconductivity. At lower laser power (Fig.\,\ref{FigSIreconstruction}(left column)), the contrast is mostly uniform again along the Nb wire, indicating that it is fully superconducting with no normal regions, and the missing current is recovered. In previous work, a similar effect of missing current was observed in normal metals at room temperature,\cite{Tetienne2019} where it was interpreted as an apparent delocalisation of the current towards the diamond substrate, which decreases $|\tilde{\bf{J}}|$ in the first method but not in the second method. A possible explanation for this effect may involve a magnetic response in the diamond itself which can be significant for nitrogen-doped diamond at low temperatures,\cite{Barzola-Quiquia2019} however this requires further investigation (see Ref.\cite{Tetienne2019} for a more complete discussion). For this reason, we chose to use the second method in the main text, assuming the resulting $|\tilde{\bf{J}}|$ is a good indicator for the transport in the Nb film. With the second method, the integrated current $I_{\rm int.}$ remains within $10$\% of the electrically measured value ($20$\,mA) in all cases at any point along the wire.    

So far we have presented results from a single location along the Nb wire. In Fig.\,\ref{FigSIJvsLocation}, we show the current density $|\tilde{\bf{J}}|$ (obtained with the second method) for two other locations [Fig.\,\ref{FigSIJvsLocation}(a,d) and (c,f)] and compare with the original location [Fig.\,\ref{FigSIJvsLocation}(b,e)]. The current pattern is broadly similar in all cases, supporting the interpretation that it is mostly dictated by the laser profile which imprints a temperature profile. Moving the sample relative to the laser does not change the temperature profile in the field of view and therefore, the current density pattern is essentially unchanged. There are small differences still, and those somewhat correlate with features seen in the ODMR contrast [Fig.\,\ref{FigSIJvsLocation}(d-f)]. For instance, at the left most location [Fig.\,\ref{FigSIJvsLocation}(a,d)] a hot spot of current can be seen which corresponds to an island of normal ODMR contrast (i.e. a fully superconducting region) in the middle of a patch of reduced ODMR contrast (i.e. a region of reduced superconductivity).

\end{document}